\documentclass[acmsmall authorversion]{acmart}

\makeatletter
\def\@ACM@checkaffil{%
    \if@ACM@instpresent\else
    \ClassWarningNoLine{\@classname}{No institution present for an affiliation}%
    \fi
    \if@ACM@citypresent\else
    \ClassWarningNoLine{\@classname}{No city present for an affiliation}%
    \fi
    \if@ACM@countrypresent\else
        \ClassWarningNoLine{\@classname}{No country present for an affiliation}%
    \fi
}
\makeatother
\AtBeginDocument{%
  }

\author{Alexandre LANVIN}
\orcid{https://orcid.org/0009-0003-5343-2528}
\affiliation{Inria, Université Côte d’Azur, France}
\email{Alexandre.Lanvin@inria.fr}

\author{JEFFREY HU}
\orcid{https://orcid.org/0009-0003-4400-0862}
\affiliation{Inria, Université Côte d’Azur, France}
\affiliation{Cambridge University, United Kingdom}
\email{hujh14@gmail.com}

\author{SIMON LUCAS}
\orcid{https://orcid.org/0009-0002-6134-7668}
\affiliation{Inria, Université Côte d’Azur, France}
\email{simon.lucas@inria.fr}

\author{ADRIEN BOUSSEAU}
\orcid{https://orcid.org/0000-0002-8003-9575}
\affiliation{Inria, Université Côte d’Azur, France}
\email{adrien.bousseau@inria.fr}

\author{GEORGE DRETTAKIS}
\orcid{https://orcid.org/0000-0002-9254-4819}
\affiliation{Inria, Université Côte d’Azur, France}
\email{George.Drettakis@inria.fr}

\acmConference[Conference I3D '26]{Make sure to enter the correct
  conference title from your rights confirmation email}{2026}{San Fransisco, CA}

\acmSubmissionID{19}

\begin{document}

\setcopyright{rightsretained}
\acmJournal{PACMCGIT}
\acmYear{2026} \acmVolume{9} \acmNumber{1} \acmArticle{10,19}
\acmMonth{5} \acmDOI{10.1145/3804495}

\title{Intrinsic decomposition and editing of 3D Gaussian splats}

\begin{teaserfigure}
  \includegraphics[width=\textwidth]{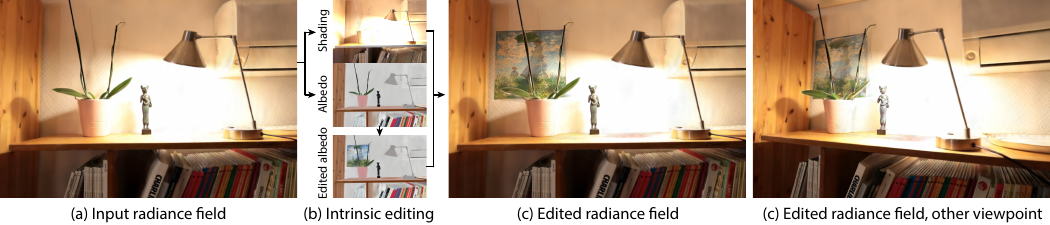}
  \caption{By decoupling albedo and shading in a radiance field (a-b) our method enables direct edition of albedo in a physically plausible way while remaining multi-view consistent (c-d).}
  \Description{The teaser figure shows an image of a room decomposed into albedo and shading. The albedo is then edited by adding a painting on a wall. Multiplying the edited albedo with the shading produces a realistic edited image, which can be seen from several viewpoints.}
  \label{fig:teaser}
\end{teaserfigure}

\begin{abstract}
  Intrinsic decomposition — which expresses image colors as the product of diffuse albedo and shading, possibly augmented with view-dependent residuals —  has a long history in image editing as it enables the modification of object colors and textures without altering lighting. We extend intrinsic decomposition to radiance fields represented with Gaussian splatting by proposing solutions to three key aspects of such decomposition. First, we describe how to model the intrinsic decomposition as independent sets of Gaussian primitives, which allows each set to adapt to the characteristics of the layer it represents. Second, we present an optimization procedure guided by data-driven predictions to disentangle multi-view photographs of a scene into the aforementioned intrinsic sets. Finally, we provide an editing workflow where users modify the texture of planar surfaces simply by modifying the albedo of that surface in one image. Capturing this edit within the intrinsic radiance field allows re-rendering of the edited scene with plausible lighting under arbitrary viewpoints.
\end{abstract}

\ccsdesc[500]{Computing methodologies~Rasterization}
\ccsdesc[300]{Computing methodologies~Neural networks}
\ccsdesc[500]{Computing methodologies~Reflectance modeling}

\maketitle

\section{Introduction}
Intrinsic decomposition is a fundamental task for image editing as it allows modification of object colors (called \emph{albedo}) without altering shading and shadows \cite{BKPB17}. Numerous algorithms have been proposed to decompose images \cite{BPD09}, image collections \cite{Laffont2012}, videos \cite{BSTSPP14}, lightfields \cite{Garces2017}, and neural radiance fields \cite{Ye2023IntrinsicNeRF}. In this paper, we describe how to \emph{model}, \emph{reconstruct} and \emph{edit} intrinsic decompositions of radiance fields represented as 3D Gaussian splats \cite{KKLD23}, allowing real-time navigation in the modified 3D scene.

\paragraph{Modeling intrinsic decompositions of Gaussian splats.}
3D Gaussian splatting (3DGS) represents the radiance field of a scene as a collection of 3D Gaussian primitives that store localized view-dependent colors, typically optimized to reproduce the colors observed in multiple photographs of the scene. 
Prior work proposed to decompose 3D Gaussian splats by augmenting each primitive with intrinsic quantities, i.e, to separate the view-dependent color of each Gaussian into its albedo, shading and view-dependent residual values (or other material-lighting separations) \cite{jiangGaussianShader3DGaussian2024,liangGSIR3DGaussian2024,ye3DGaussianSplatting2024}. 
But while intrinsic quantities often exhibit some spatial correlation (for example, object boundaries typically produce discontinuities in both albedo and shading), they can also require significantly different spatial resolution (for example, object textures produce high-frequency details in albedo but not in shading, while hard shadows produce sharp edges in shading that should not appear in albedo). We thus propose to model and reconstruct intrinsic decompositions with \emph{separate sets of Gaussians} for the albedo, shading and residual components, allowing our approach to adapt the number and distribution of Gaussian primitives to the local characteristics of each signal.

\paragraph{Reconstructing intrinsic decompositions of Gaussian splats.}
Decomposing observed radiance into intrinsic quantities is an ill-posed problem, for which numerous heuristic \cite{BKPB17} and data-driven \cite{garces2022surveyintrinsic} priors have been proposed. We follow the latter approach and leverage a recent diffusion-based video decomposition model \cite{DiffusionRenderer} to predict the albedo component of all images of a scene. Compared to single-image models, this video model produces consistent predictions across successive images, which is critical to fuse these predictions into a single 3D Gaussian splat.
However, being free of shading, albedo images exhibit large uniform areas that hinder 3D reconstruction. We address this challenge by regularizing the 3D reconstruction of albedo Gaussians with per-image depth predictions computed on the original input images. Equipped with a consistent \emph{albedo field}, we next optimize a view-independent \emph{shading field} and a view-dependent \emph{residual field} -- each represented by its own set of Gaussians -- such that rendering and compositing all three fields reproduces the input images.

\begin{figure}[!t]
\centering
  \includegraphics[width=\textwidth]{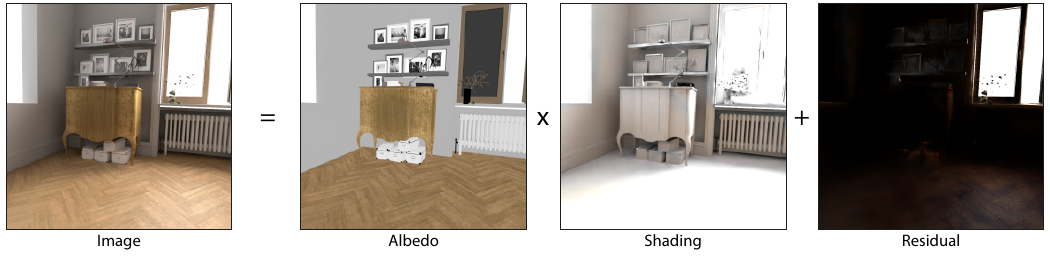}
  \caption{The intrinsic image model expresses the observed image (left) as the product of diffuse albedo and shading, summed with a view-dependent residual (right).}
  \Description{Illustration of our image formation model on a typical image}
  \label{fig_image_formation_model_diffuse}
\end{figure}

\paragraph{Editing intrinsic decompositions of Gaussian splats.}
Similar to prior work on intrinsic images, we demonstrate the value of our decomposition by modifying object colors and texture while preserving realistic shading and shadows (Figure~\ref{fig:teaser}). With our solution, users can orient a virtual camera towards the surface they wish to edit, and paint over the resulting image to modify its albedo. We then update the 3D Gaussians that represent the albedo field to capture these user edits while keeping the shading Gaussians untouched. The main difficulty resides in ensuring that the edits performed in one view stay consistent in other views, which is not guaranteed when a single view is used for Gaussian optimization. To address this difficulty, we reconstruct a proxy surface around the user edits and use it to reproject the edits in nearby views to further constrain the albedo update.

In summary, our contributions are:
\begin{itemize}
    \item A representation of radiance fields as three separate Gaussian splats representing albedo, shading and view-dependent residuals.
    \item A method to reconstruct each of these intrinsic fields from multiple photographs, based on a video diffusion prior.
    \item A method that leverages this intrinsic decomposition to edit the albedo of a scene while keeping the shading untouched, allowing free-viewpoint navigation in the edited scene.
\end{itemize}

We demonstrate our approach by decomposing and editing radiance field reconstructions of three real-world scenes, and of one synthetic scene for which we have ground truth albedo.

\section{Related Work}
There is a vast literature on intrinsic image decomposition, typically separating the image into diffuse albedo and lighting layers. In later work, a residual layer was also introduced to better model view-dependent effects. We refer the reader to the excellent surveys covering this body of work~\cite{BKPB17,garces2022surveyintrinsic}. Early approaches depended on heuristic priors, for example assuming that the albedo is piecewise-constant while lighting is smooth, with mixed results. Recent approaches employ machine learning to obtain more powerful, data-driven priors~\cite{careaga2023intrinsic,careaga2024colorful}.
In particular, diffusion models offer rich priors learned from very large datasets, and recent work has demonstrated that fine-tuning a large diffusion model can achieve impressive intrinsic decompositions~\cite{RGB2X2024Zeng,kocsis2024iid,IntrinsicDiffusion2024Luo}. Since we are interested in the multi-view context, we employ a video-based diffusion model~\cite{DiffusionRenderer} to extract coherent intrinsic layers for multiple, densely-captured  photographs of a scene.

Intrinsic decomposition is often a key component for relighting captured scenes. Early work combined forward rendering with heuristic priors~\cite{DRCLLPD15}, while subsequent methods adopted deep learning for better results~\cite{PMGD21}. The introduction of Neural Radiance Fields (NeRFs) quickly led to exploration of how to jointly perform 3D reconstruction, intrinsic decomposition and relighting~\cite{Ye2023IntrinsicNeRF,boss2021nerd,boss2021neuralpil,nerv2021,NeRFactor}, but the computational and memory expense of the original method plus the additional terms is prohibitive for all practical purposes. 3D Gaussian Splatting (3DGS)~\cite{KKLD23} reduces the computational cost, and a large number of methods consider intrinsic decompositions of Gaussian splats. However, most of these (e.g.,~\cite{Du_2025_ICCV,bai2025garerelightable3dgaussian,jiangGaussianShader3DGaussian2024,liangGSIR3DGaussian2024,ye3DGaussianSplatting2024}) perform some form of inverse rendering under environment map lighting -- often applied to an isolated object -- and thus have difficulty handling the more involved context of full scenes with local lighting. We focus on such scenes and avoid the need for full inverse rendering by relying instead on a pre-trained diffusion model for intrinsic decomposition. While our decomposition does not allow relighting, it does support convincing albedo edits.

Several methods have been develop for editing NeRFs, and in particular modifying the color, e.g., for posterization~\cite{tojo2022posternerf} or palette-based color editing~\cite{Kuang_2023_CVPR}. The speed of 3DGS has allowed the development of interactive editing tools for radiance fields, such as SplatShop~\cite{schutz2025splatshop}, that allow editing of the geometry and ``painting'' on top of existing splats, with an emphasis on Virtual Reality interaction. ReCoGS~\cite{rutayisire2025recogs} allows real-time re-coloring of splats using unprojection and fine-tuning to obtain the final result. With our intrinsic decomposition of 3DGS we can go much further than simple recoloring, since by separating the lighting from the albedo/texture we can perfom color and texture edits while preserving the lighting conditions in the scene.

\section{Method}

\begin{figure}[!t]
\centering
  \includegraphics[width=\textwidth]{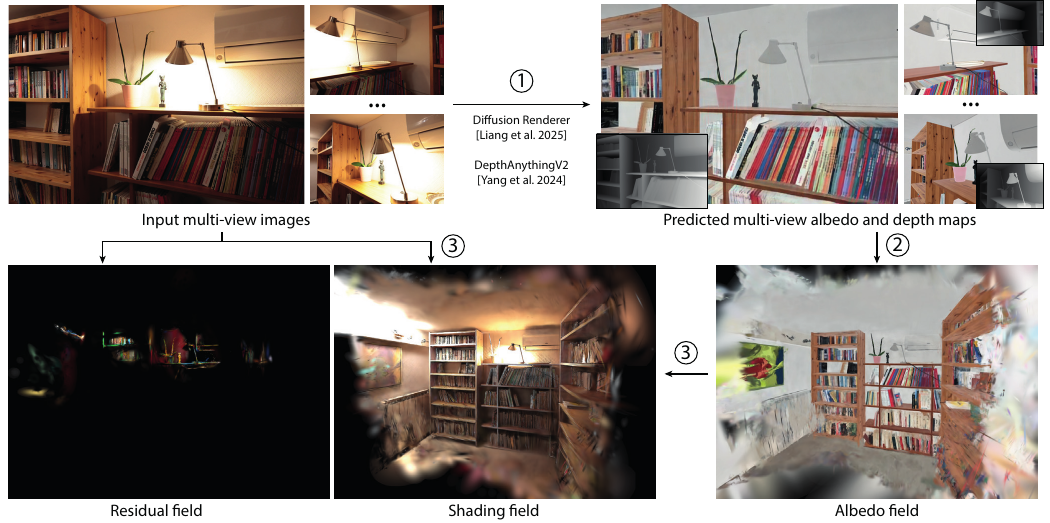}
  \caption{Overview of the pipeline for intrinsic reconstruction using distincts sets of Gaussians. Our pipeline takes as input multiple photographs of a scene and uses off the shelf methods to predict depth maps and albedo maps for each photograph (1). We first use these albedo and depth maps to reconstruct an albedo field (2, Sec.\ref{albedo_reconstruction}), and then optimize a shading field and a view-dependent residual field such that they best reproduce the input images when combined with the albedo (3, Sec.\ref{shading_reconstruction}). Users can edit the albedo by placing textured planes in the scene, our method updates the underlying albedo field accordingly (Sec.\ref{sec:albedo_editing}).
   }
  \Description{Illustration of the main steps of our method on a few images of a room.}
  \label{fig:overview_decomp}
\end{figure}

Our pipeline is illustrated in Fig.~\ref{fig:overview_decomp}. We capture a scene with a video, which we feed to a video diffusion model to estimate albedo for each frame~\cite{DiffusionRenderer}. 
We also feed the video to a monocular depth estimator to obtain an initial guess of the scene geometry~\cite{depth_anything_v2}.
We then subsample the video to create a typical multi-view dataset used to create a radiance field (a few hundred images in most of our examples). %
Equipped with this multi-view dataset and accompanying albedo estimates, the second step of our method is to create an \emph{albedo field} by optimizing a 3DGS representation from the albedo images. We regularize the optimization using the estimated depth. In the third step, we optimize a \emph{shading field} and a \emph{residual field} that we composite with the albedo field to reproduce the original input images. With this representation, the user can directly edit the albedo layer, e.g., by painting or by adding a texture. We then re-optimize the albedo Gaussians to capture the modified texture, allowing us to render the modified scene from any viewpoint while maintaining the original lighting effects. We next describe each step in detail.

\subsection{Modeling intrinsic decompositions of Gaussian splats}
We adopt the intrinsic image formation model that expresses each observed image $I$ as the product of a diffuse albedo image $A$ and a diffuse shading image $S$, summed with a view-dependent residual $R$ \cite{Ye2023IntrinsicNeRF,careaga2024colorful} (Fig.~\ref{fig_image_formation_model_diffuse}):
\begin{equation}
\label{eq:basic_im}
I = A \times S + R.
\end{equation}
Prior work on intrinsic decomposition of Gaussian splats \cite{jiangGaussianShader3DGaussian2024,liangGSIR3DGaussian2024,ye3DGaussianSplatting2024} (resp. neural radiance fields \cite{Ye2023IntrinsicNeRF}) optimize a single set of Gaussians (resp. a single MLP) to express all intrinsic quantities. Instead, we propose to optimize three separate sets of Gaussians $\mathcal{G}_{A}$, $\mathcal{G}_{S}$ and $\mathcal{G}_{R}$ which, once splatted and composited according to Eq.~\ref{eq:basic_im}, reproduce all input images. 
Using three separate sets allows our approach to better capture the (possibly misaligned) sharp details of each intrinsic component.

Similarly to standard 3DGS \cite{KKLD23}, we model each Gaussian primitive within $\mathcal{G}_{A}$, $\mathcal{G}_{S}$ or $\mathcal{G}_{R}$ with a position $p$, color $c$, opacity $\alpha$ and a covariance matrix $\Sigma$:
\begin{equation}
    g = \{c, \alpha, p,\Sigma\}.
\end{equation}
However, while the original method uses spherical harmonics to encode view-dependent radiance in $c$, we only use spherical harmonics for the Gaussians $\mathcal{G}_{R}$ that model the view-dependent residual, and resort to a simple triplet $c = \{r,g,b\} \in \mathbb{R}^3$ for the Gaussians $\mathcal{G}_{A}$ and Gaussians $\mathcal{G}_{S}$ that model diffuse albedo and shading, respectively.
For a given viewpoint, we render each set of Gaussians into its corresponding image by using the same rasterizer as in the original 3DGS implementation. 
The three sets of Gaussians $\mathcal{G}_{A}$, $\mathcal{G}_{S}$ and $\mathcal{G}_{R}$ can be seen as separate 3D fields, which we call \emph{albedo}, \emph{shading} and \emph{residual fields}.

\subsection{Reconstructing intrinsic decompositions of Gaussian splats}
We now describe how to reconstruct the three sets of intrinsic Gaussians $\mathcal{G}_{A}$, $\mathcal{G}_{S}$ and $\mathcal{G}_{R}$ from multiple photographs of a scene. Given the ambiguity of the task, we first leverage a pre-trained video diffusion model to estimate the albedo for each input image. This data-driven estimate allows us to reconstruct $\mathcal{G}_{A}$, which we then freeze to optimize $\mathcal{G}_{S}$ and $\mathcal{G}_{R}$ to best reconstruct the images according to Eq.~\ref{eq:basic_im}.

\subsubsection{Data-driven albedo estimation}
To extract albedo for each input image, we employ DiffusionRenderer, a state-of-the-art material estimation method based on video diffusion models~\cite{liang2025diffusion}, which yields temporally consistent albedo estimates across all frames. Our initial experiments with single-image methods (e.g.,~\cite{RGB2X2024Zeng}) revealed significant inter-frame inconsistencies that degrade 3D reconstruction quality.

In practice, video capture produces sequences containing hundreds of frames, whereas DiffusionRenderer is limited to processing clips of 57 frames at 24 FPS. Naïvely partitioning a long video into independent chunks and denoising each chunk separately introduces visible discontinuities in albedo predictions at chunk boundaries. To mitigate this issue, we adopt Brick Diffusion~\cite{yuan2025brick} and stagger the denoising process across neighboring chunks, enabling smooth transitions between the chunks.

Cosmos DiffusionRenderer operates in the latent space of the Cosmos Video Tokenizer~\cite{agarwal2025cosmos}. However, the tokenizer’s asymmetric design complicates the direct application of brick diffusion. DiffusionRenderer denoises eight latent codes per step, where the first latent decodes to a single frame, and each subsequent latent decodes to a group of eight frames, yielding a total of 57 frames. Effective application of brick diffusion requires shifting the denoising window by both an integer number of latent codes and an integer number of frames simultaneously.

To address this, we process the sequence in chunks of 64 frames, corresponding to one start latent and eight regular latents. At each denoising step, we operate on 57 out of the 64 frames to denoise one start latent and seven of the eight regular latents. The window is then advanced by four latent codes, corresponding to a shift of 32 frames. After denoising, all start and regular latents are decoded to frames, and redundant frames are discarded. This strategy produces temporally consistent albedo predictions over long sequences with minimal boundary artifacts.

The video diffusion model expects and produces video at 24 FPS, which is far too dense for radiance field reconstruction. Hence the set of input images we use in our method is a subset of the video frames, it is obtained by subsampling the video sequence to a few hundred images taking every $n$th frame, with a preference for frames with low blur. We determined experimentally that sampling every $n=5$ frame provided a sufficiently dense camera path for high quality novel view synthesis. We then run Structure-from-Motion (SfM)~\cite{schoenberger2016sfm} on the extracted images to compute camera poses. Importantly, we apply the undistortion SfM applies on the input images to the albedo images as well.

\subsubsection{Albedo reconstruction}\label{albedo_reconstruction}
As can be seen in Fig.~\ref{fig_image_formation_model_diffuse}, the albedo images typically contain piecewise constant color regions with an insufficient number of geometric cues, making it harder for the 3DGS algorithm to place Gaussians in 3D space. We address this challenge by regularizing the reconstruction against depth maps estimated for each input view \cite{chung2024depth}, which we obtain using the Depth Anything V2 monocular depth estimator \cite{depth_anything_v2}.

As in 3DGS, we start from a set of calibrated cameras and an initial point cloud typically produced by SfM. The set of albedo Gaussians $\mathcal{G}_{A}$ is optimized using the same learning rates and activation functions as in the original implementation except for the albedo color, which we expect to be in $[0;1]$ so we use a sigmoid activation function. 
For regularization, we follow \cite{chung2024depth} and render a depth image $D$ by replacing the color of the Gaussians by their depth. Given the predicted depthmap $\tilde{D}$, we define the regularization term as:
\begin{equation}
    \mathcal{L}_{depth}=||\tilde{D} - D||_1.
\end{equation}
This regularization term encourages the placement of Gaussians close to the actual geometry of the scene and greatly improves densification in texture-free regions, such as the constant color patches frequently observed in the input albedo images. Depth regularization also helps snapping the user edits to the scene surface during albedo editing (Sec.\ref{sec:albedo_editing}).

To improve multi-view consistency, we add another regularization term on the average scale $\bar{s}$ of each Gaussian primitive to encourage the primitives to flatten and form sharp surface boundaries:
\begin{equation}
    \mathcal{L}_{scale}= \bar{s}.
\end{equation}

Considering the predicted albedo images $\tilde{A}$ and rendered albedo images $A$, the total loss function for reconstructing the albedo radiance field is computed as:
\begin{equation}\label{albedo_loss}
    \mathcal{L} = (1 - \lambda) ||\tilde{A} - A||_1 + \lambda \mathcal{L}_{D-SSIM}(\tilde{A},A) + \lambda_d \mathcal{L}_{depth} + \lambda_s\mathcal{L}_{scale}.
\end{equation}

An important benefit of our approach is that the albedo field is decoupled from shading and thus does not reconstruct lighting variations and shadows. This results in more uniform distribution of Gaussian primitives across the scene, which in turn allows more efficient editing of albedo and texture.

\subsubsection{Shading and Residual Reconstruction}\label{shading_reconstruction}
Once the albedo field is reconstructed, the $\mathcal{G}_{A}$ splats can be rendered from any viewpoint to produce albedo images. Given an input image $I$ and the rendered albedo image $A$, we invert the image formation model (Eq.\ref{eq:basic_im}) to optimize the shading and residual Gaussians $\mathcal{G}_{S}$ and $\mathcal{G}_{R}$. Denoting $S$ and $R$ the rendered shading and residual, we aim to minimize the reconstruction loss:
\begin{equation}
  \mathcal{L} = (1 - \lambda) ||I - (A \times S + R)||_1 + \mathcal{L}_{D-SSIM}(I , A \times S + R) + \lambda_d \mathcal{L}_{depth} + \lambda_s\mathcal{L}_{scale}
  \label{eq:full_reconstruction}
\end{equation}

However, since jointly optimizing $\mathcal{G}_{S}$ and $\mathcal{G}_{R}$ is a challenging task, we perform this optimization in two phases. First, we ignore the residual term and only optimize $\mathcal{G}_{S}$ using the diffuse image formation model $I = A \times S$; these
shading Gaussians are initialized from SfM points in the same manner as albedo Gaussians.
Once this first optimization is completed, we freeze $\mathcal{G}_{S}$ and optimize $\mathcal{G}_{R}$ to capture residual non-diffuse effects.

Furthermore, since the shading of a scene is unbounded and can exhibit high dynamic range, we follow the recommendations of \citet{careaga2023intrinsic} and compute the photometric loss in the inverse domain, i.e., minimizing the difference between $\frac{1}{1 + I}$ and $\frac{1}{1 + A \times S + R}$. This transformation bounds the values in $[0;1]$, alleviating the need for an activation function on the color parameters.

If an albedo channel is dark, the shading can become arbitrarily large, which may lead to incorrect color reproduction.
To alleviate this problem, we encourage the shading to be gray with an additional regularization term that computes, for each shading Gaussian, the range of its color parameters:
\begin{equation}
\label{eq:col_reg}
\mathcal{L}_{gray} = max(c) - min(c).
\end{equation}

Residual Gaussians $\mathcal{G}_{R}$ are created where they are most likely to be needed. Specifically, we iterate through the training views and compute the difference between the input image and the composited diffuse render (i.e., $A \times S)$. 
We randomly select a subset of pixels with high difference and use their predicted depth as positions to spawn residual Gaussians.
We model the color of these residual Gaussians with up to 3 bands of spherical harmonics to represent view-dependent effects. We use the same scheduling as in 3DGS to progressively add additional bands to the spherical harmonics and we keep the same scaling and depth regularizations as in albedo and shading reconstruction stages.
Optimization then proceeds with the full image formation model (Eq.~\ref{eq:full_reconstruction}) but only the residual Gaussians are updated.

\begin{figure}[!h]
\centering
\begin{tabular}{ccccc}
   \hspace{-3mm}\includegraphics[width=0.19\linewidth]{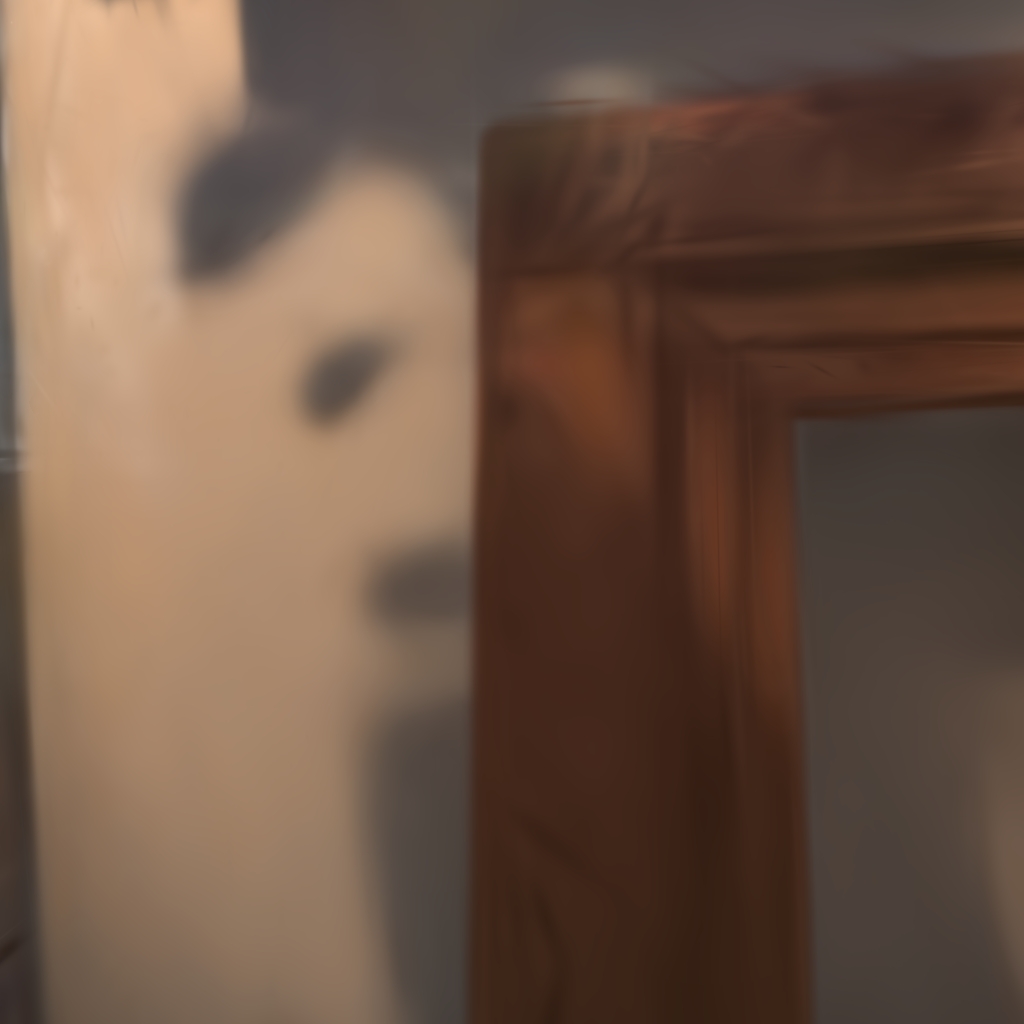} &
   \hspace{-3mm}\includegraphics[width=0.19\linewidth]{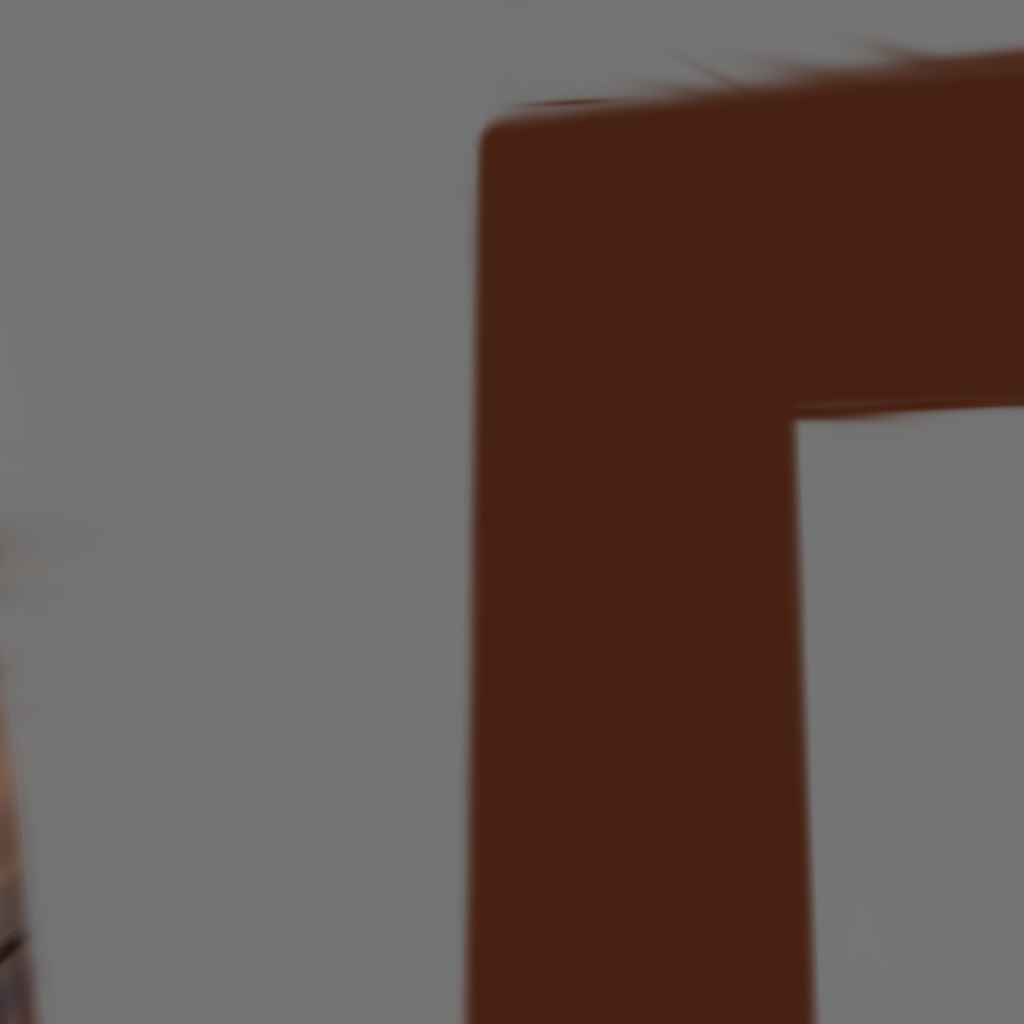} &
   \hspace{-3mm}\includegraphics[width=0.19\linewidth]{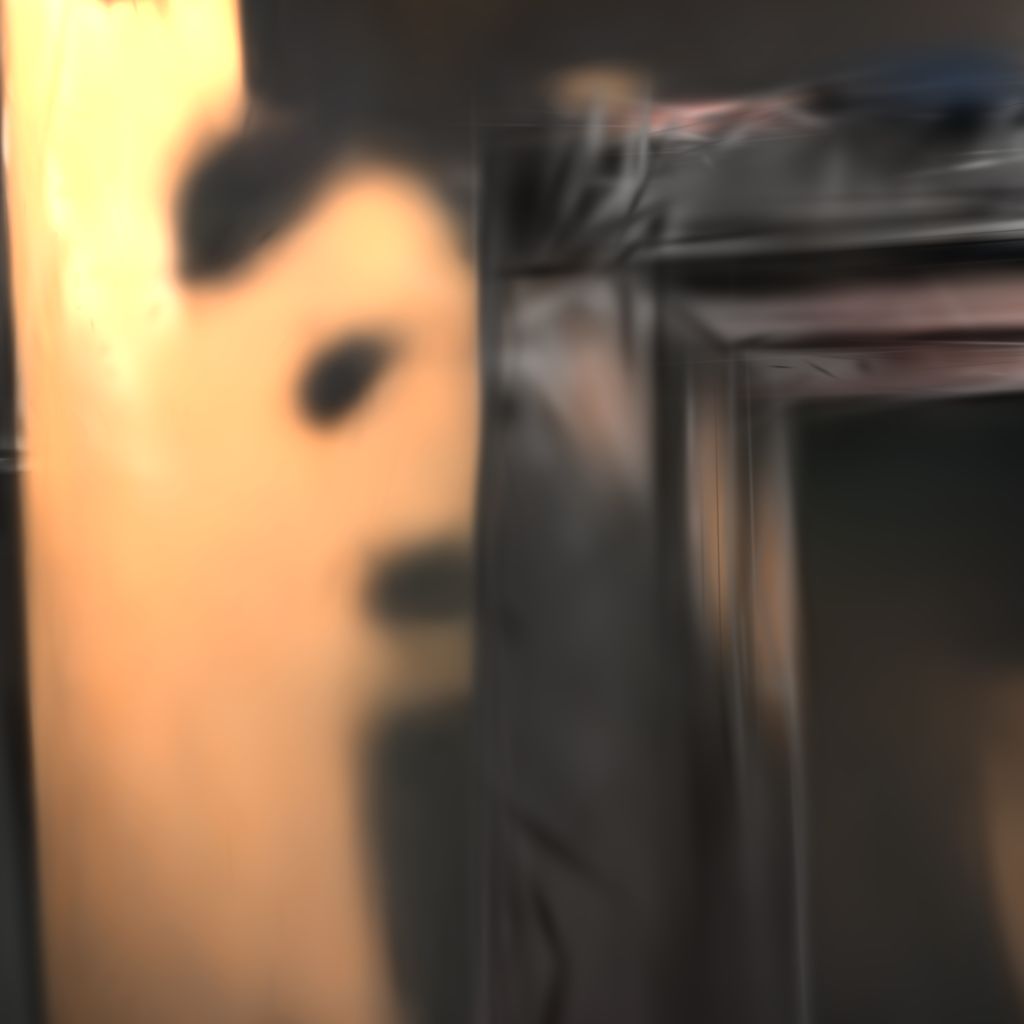} &
   \hspace{-3mm}\includegraphics[width=0.19\linewidth]{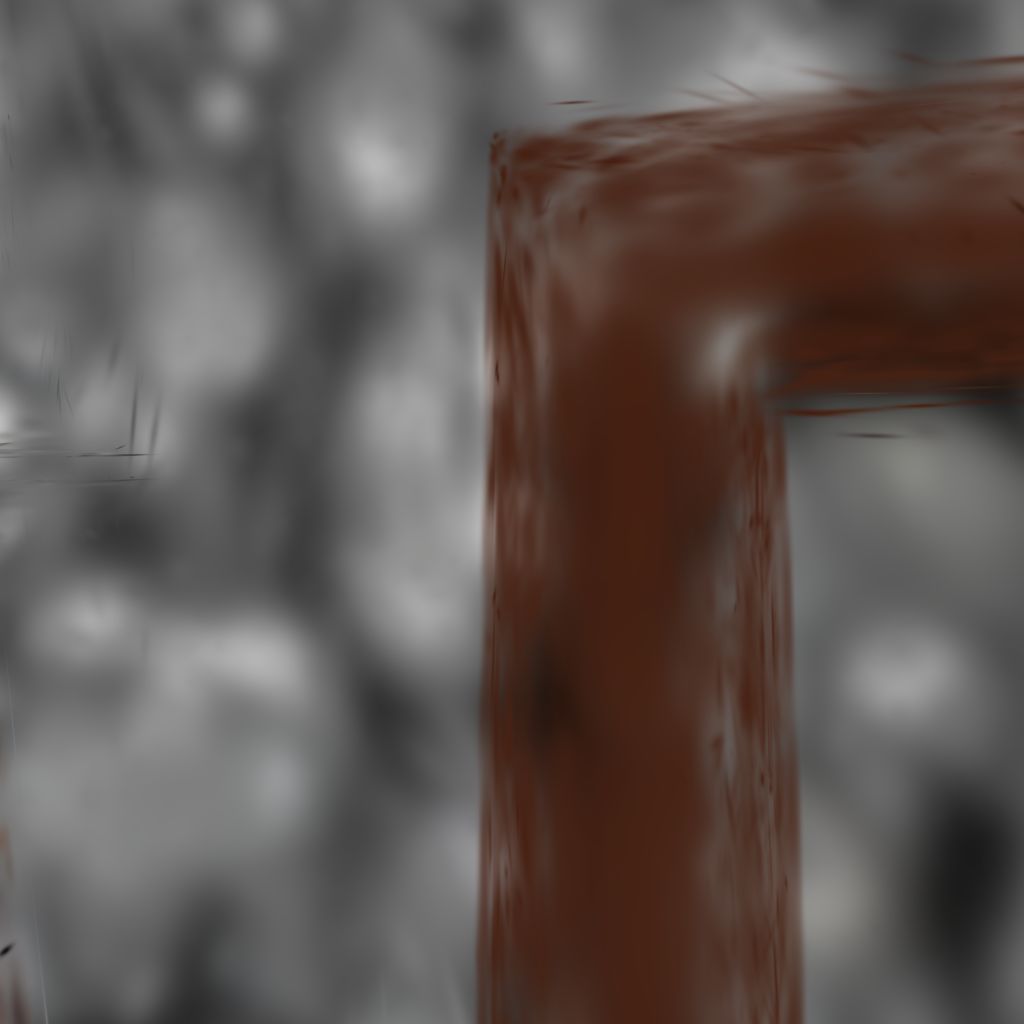} &
   \hspace{-3mm}\includegraphics[width=0.19\linewidth]{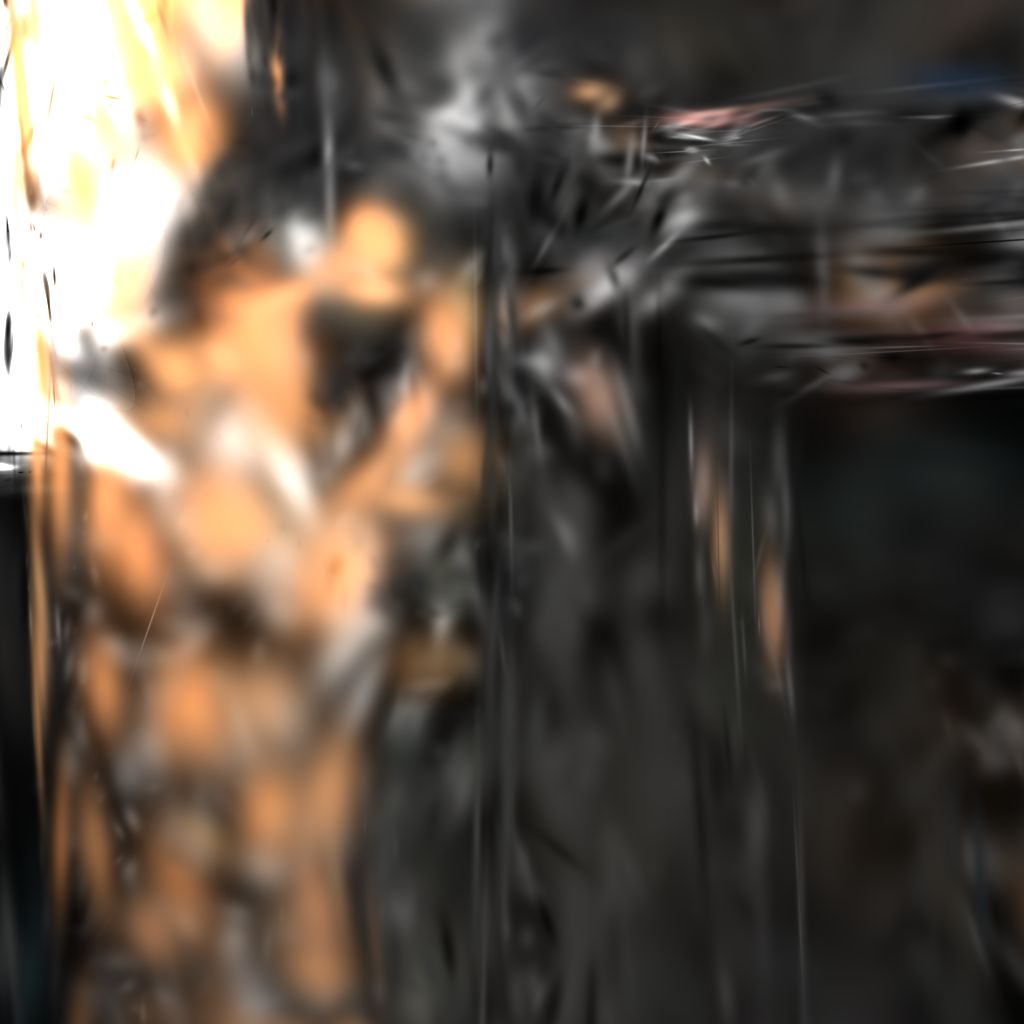} \\
   \hspace{-3mm}\footnotesize{Full Render} & \hspace{-3mm}\footnotesize{Albedo field} & \hspace{-3mm}\footnotesize{Shading field} & \hspace{-3mm}\footnotesize{Albedo field*} & \hspace{-3mm}\footnotesize{Shading field*}  
\end{tabular}
\caption{Each intrinsic field is optimized to represent different quantities. In Albedo field* and Shading field*, we have artificially scaled down the scale of the Gaussians. Notice the different distribution of primitives across albedo and shading field, each capturing details specific to each.}
\label{fig_distribution}
\end{figure}

Each optimized field is composed of primitives with positions and shapes adapted to the content of each intrinsic layer. This can be seen  in Fig.~\ref{fig_distribution} where in the two rightmost images we have artificially reduced the scale of the Gaussians to better visualize the individual primitives. We can see that small and anisotropic Gaussian primitives are created in the shading layer to model shadows, and equivalently to model texture content in albedo.

\subsection{Editing intrinsic decompositions of Gaussian splats}
\label{sec:albedo_editing}

Once we have the full radiance field decomposed in three components, it is easy to edit the albedo separately while maintaining lighting and shadow effects. We provide an interactive interface allowing the user to perform albedo editing. To better comprehend the editing process, please see the supplemental video.

Users of our interface can edit the albedo in any input view. However, while we can minimize Eq.~\ref{albedo_loss} on this view to update the albedo Gaussians accordingly, the resulting albedo field often degrades as soon as we move away from the view under which the editing was performed. Achieving multi-view consistency requires minimizing Eq.~\ref{albedo_loss} over multiple edited views, which would be prohibitive to provide by users. Our solution consists in \emph{synthesizing} these multiple views using depth-based reprojection of the user edits. 

Unfortunately, 3DGS does not provide sufficiently reliable depth to perform such reprojection into neighboring views; for this reason we use a proxy plane to approximate the surface to be edited. The user first indicates the surface of interest by dragging a rectangular window in the image. We select, for each pixel in that window, the primitive that contributes most to its rendering. We then use RANSAC to fit the proxy plane over the center of these Gaussians.
The plane can be optionally manipulated using standard 3D widgets, allowing the user to adjust its placement in the scene. Since a typical editing session involves adding or replacing the texture of a surface; we render the new texture over the proxy plane for pre-visualization.

\begin{figure*}[!h]
  \centering
  \includegraphics[width=\linewidth]{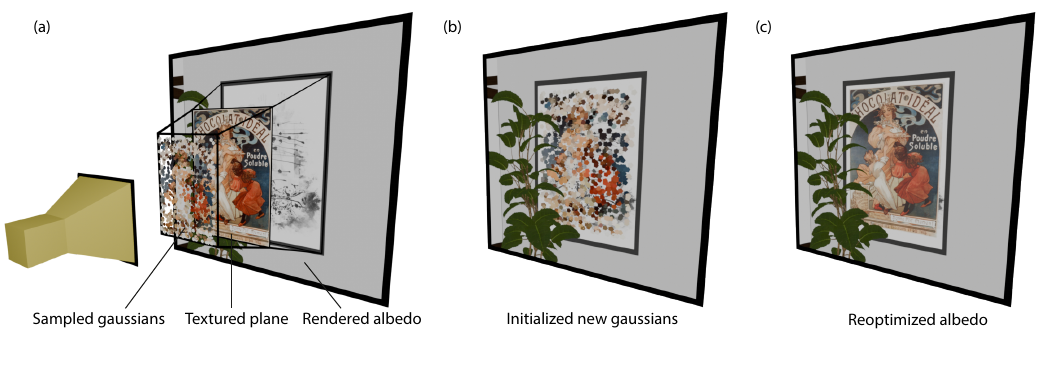}
  \caption{To capture a user edit, we initialize new Gaussians on the textured proxy plane (a). These Gaussians are added to the set of albedo Gaussians and placed in 3d space using the proxy depth (b). The augmented set of albedo Gaussians is re-optimized to match the edited albedo images rendered from multiple viewpoints (c).}
  \label{fig_albedo_edition}
\end{figure*}

Once the user is satisfied with the position of the textured proxy plane, we render a set of edited albedo images from multiple nearby viewpoints by compositing the plane with the albedo field, using the rendered depth to test for occlusions. We sample these virtual viewpoints on an hemisphere centered on the editing viewpoint.
Finally, we use these multiple edited images to re-optimize the albedo Gaussians using Eq.~\ref{albedo_loss}.
Since the user edits might contain details not present in the original albedo, 
we spawn additional albedo Gaussians on the textured proxy plane by randomly sampling $20\%$ of the pixels covered by the edit in the edited view. 
This editing process is illustrated in Fig.~\ref{fig_albedo_edition}.

\section{Implementation}

Since we are performing lighting computation, we need to perform all our compositing operations in linear space. We thus apply a simple inverse gamma correction to the sRGB images used as input to obtain ``pseudo-linear'' RGB images used for processing. We apply the same gamma inversion to the albedo maps produced by DiffusionRenderer, which we observed to be non-linear.
In addition, the albedo images can be quite blurry, resulting in floaters. We apply a simple floater reduction approach to improve our results. Our floater removal operates on each input view: We visit each pixel and accumulate all Gaussians until transmittance is saturated (``expected termination'' depth). We then compute the standard deviation of the depths of these primitives and increment a counter for those that have depth less than $n$ standard deviations of the average depth; larger floaters typically satisfy this condition for several pixels, resulting in a high value of the counter. We then remove primitives with a high counter value. This operation is performed before every densification step, which happens every 100 iterations from iteration 500 to iteration 15000 of the optimization.

\begin{figure}
    \centering
    \includegraphics[width=\linewidth]{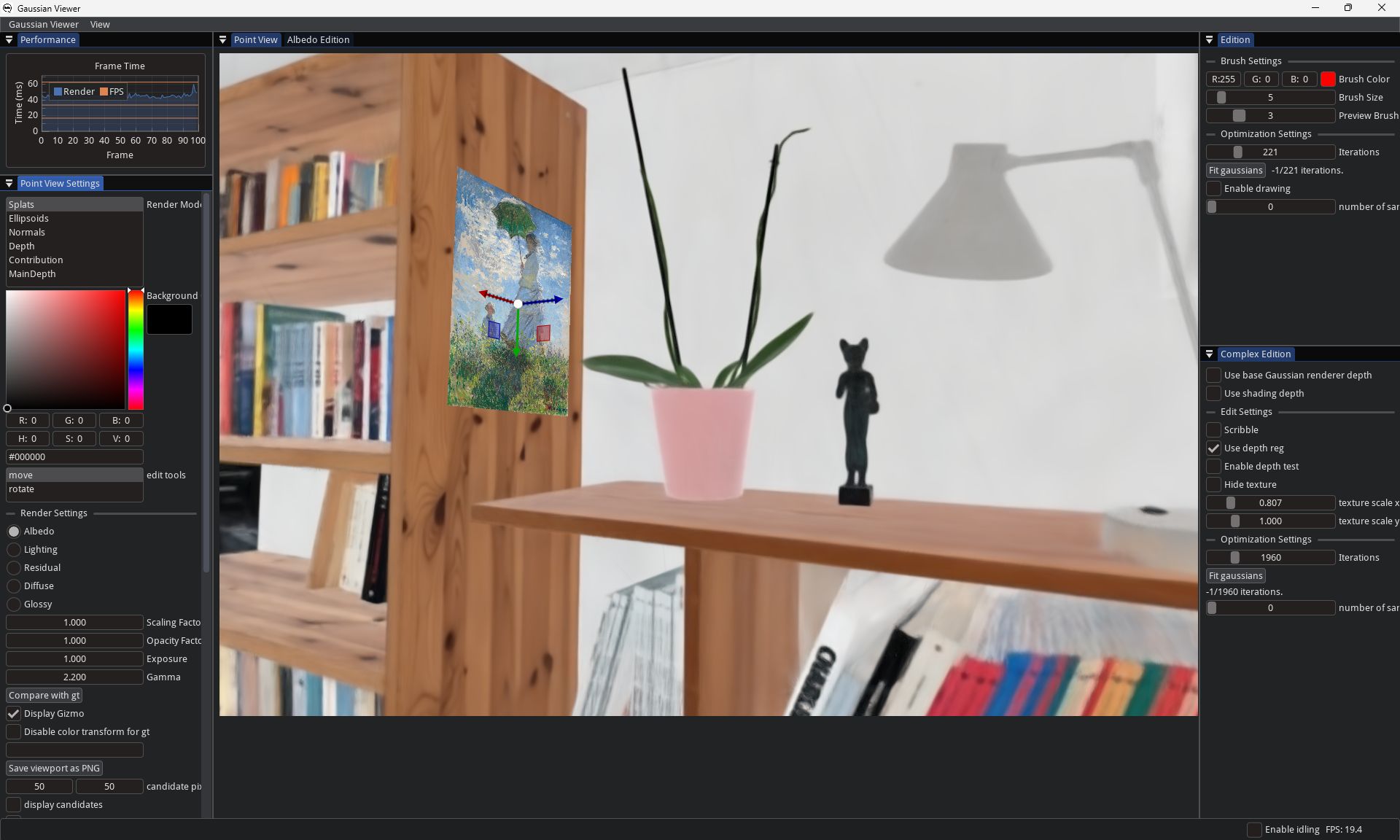}
    \caption{We added custom widgets to enable the selection of different intrinsic fields and editing of the albedo field. The user can then upload and manipulate any image in an interactive session as well as navigating the 3D reconstruction in real time.}
    \label{fig:viewer}
\end{figure}

We have implemented our system in an interactive viewer~\cite{shah_graphdecoviewer} shown in Fig.~\ref{fig:viewer}. The viewer includes interactive editing tools to select albedo Gaussians to be edited, and allows easy manipulation of the proxy plane to achieve the best results, including 3D widgets (please see supplemental video). We will release all source code of our system on publication.

\section{Results and Evaluation}
\label{sec:results}

\begin{figure*}[!h]
\centering
\includegraphics[width=\linewidth]{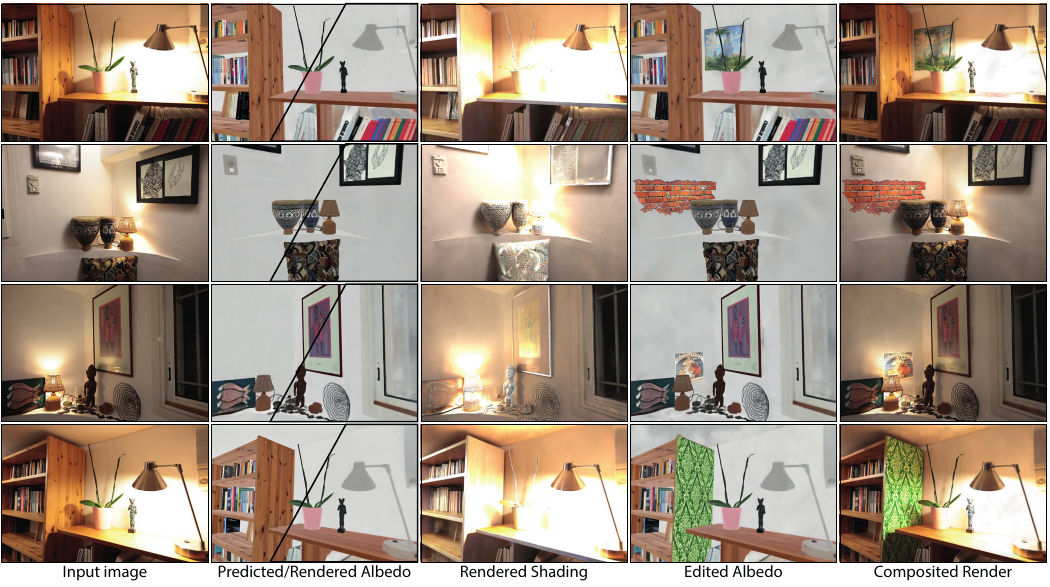}

\caption{
Intrinsic decomposition and albedo editing on real scenes. See accompanying video for animated viewpoints.
\label{fig:results}
}
\end{figure*}

In Fig.~\ref{fig:results} we show edits on three real scenes captured with a Canon EOS R6 in continuous shooting mode. The resulting sequence of still frames is dense enough to be processed as a video by DiffusionRenderer, which we use to extract albedo from the input images. Table.~\ref{tab:comparison_complexity} provides a quantitative comparison on training times and primitive counts between our method and vanilla 3DGS~\cite{KKLD23} across our 4 scenes, all experiments ran on a H100 GPU. Our method requires an additional render call every time a new layer is reconstructed (i.e., two render calls when reconstructing a diffuse scene, three when enabling the residual layer) which yields slower rendering times that inevitably leads to longer training times.

\begin{figure}[!h]
\centering
\begin{tabular}{cccc}
    \hspace{-3mm}\includegraphics[width=0.25\linewidth]{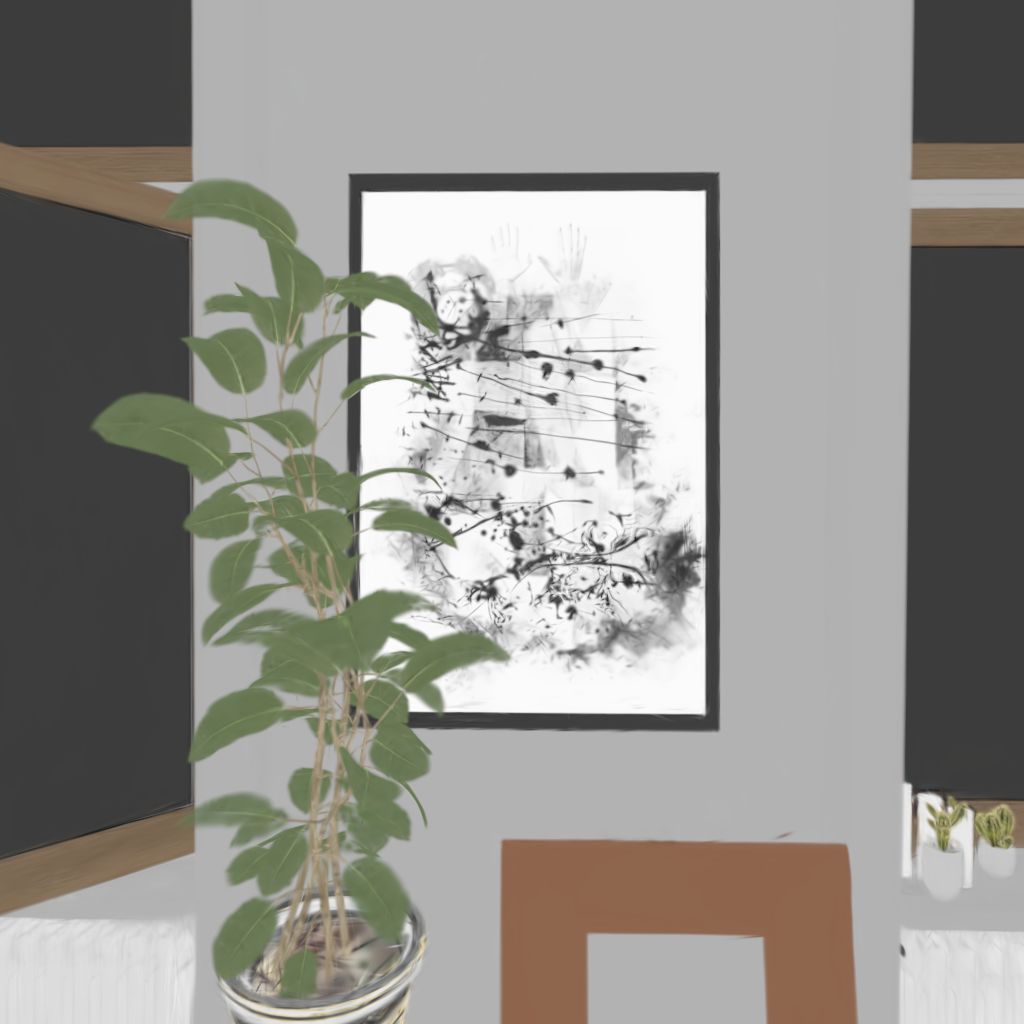} &
    \hspace{-3mm}\includegraphics[width=0.25\linewidth]{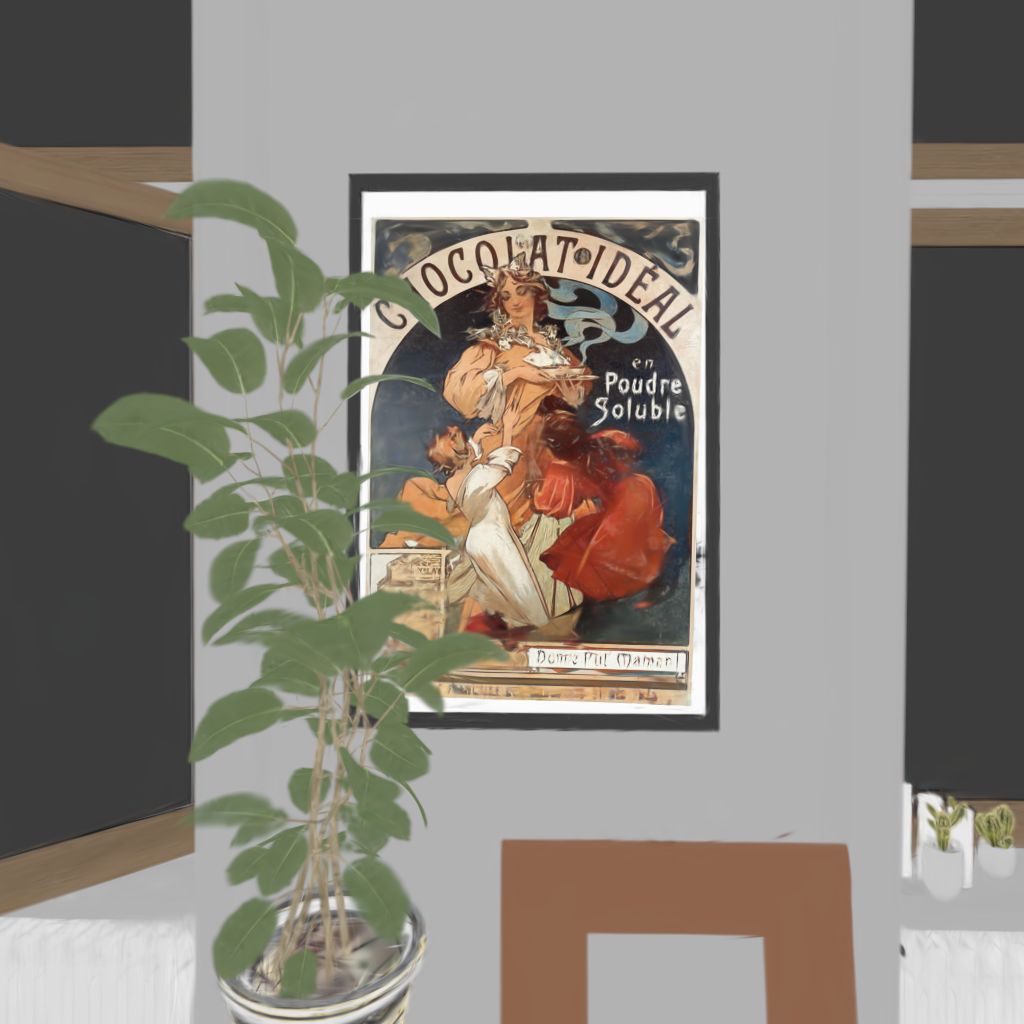} &
    \hspace{-3mm}\includegraphics[width=0.25\linewidth]{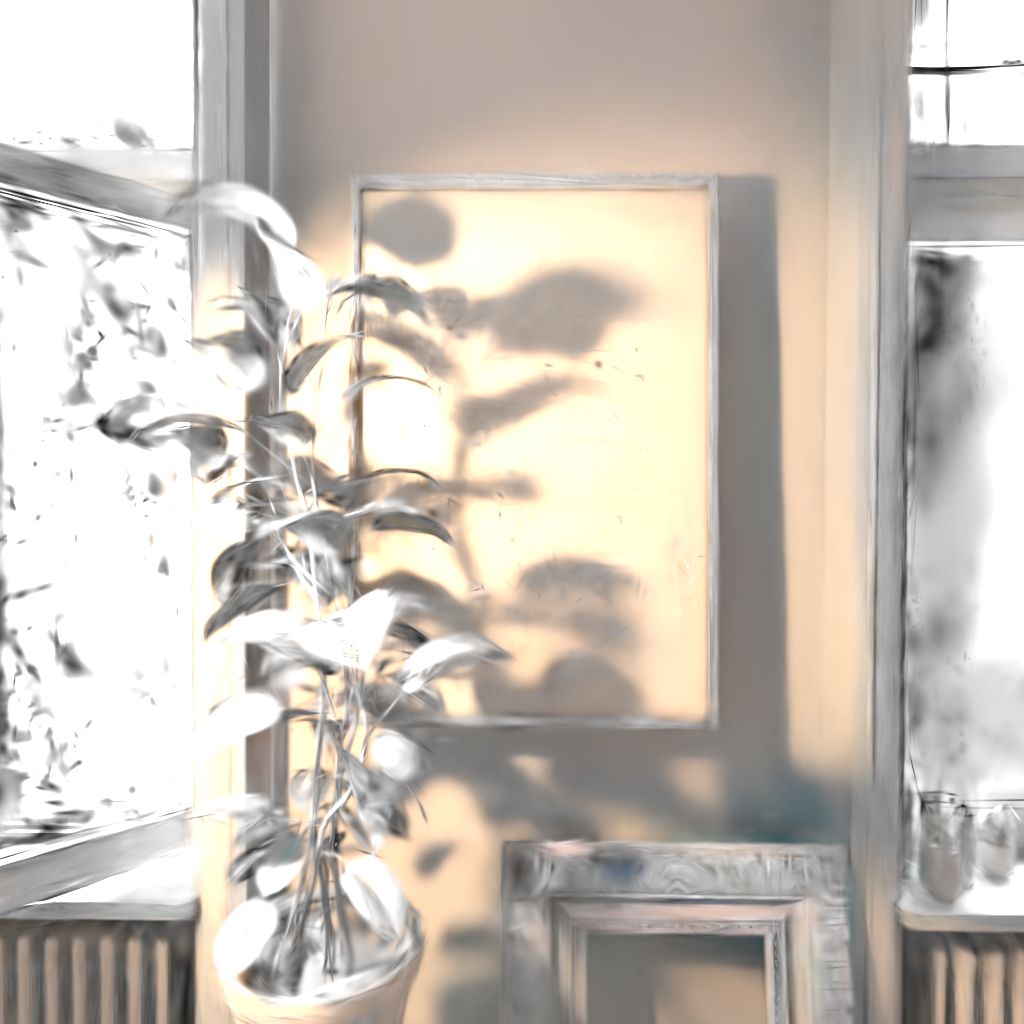} &
    \hspace{-3mm}\includegraphics[width=0.25\linewidth]{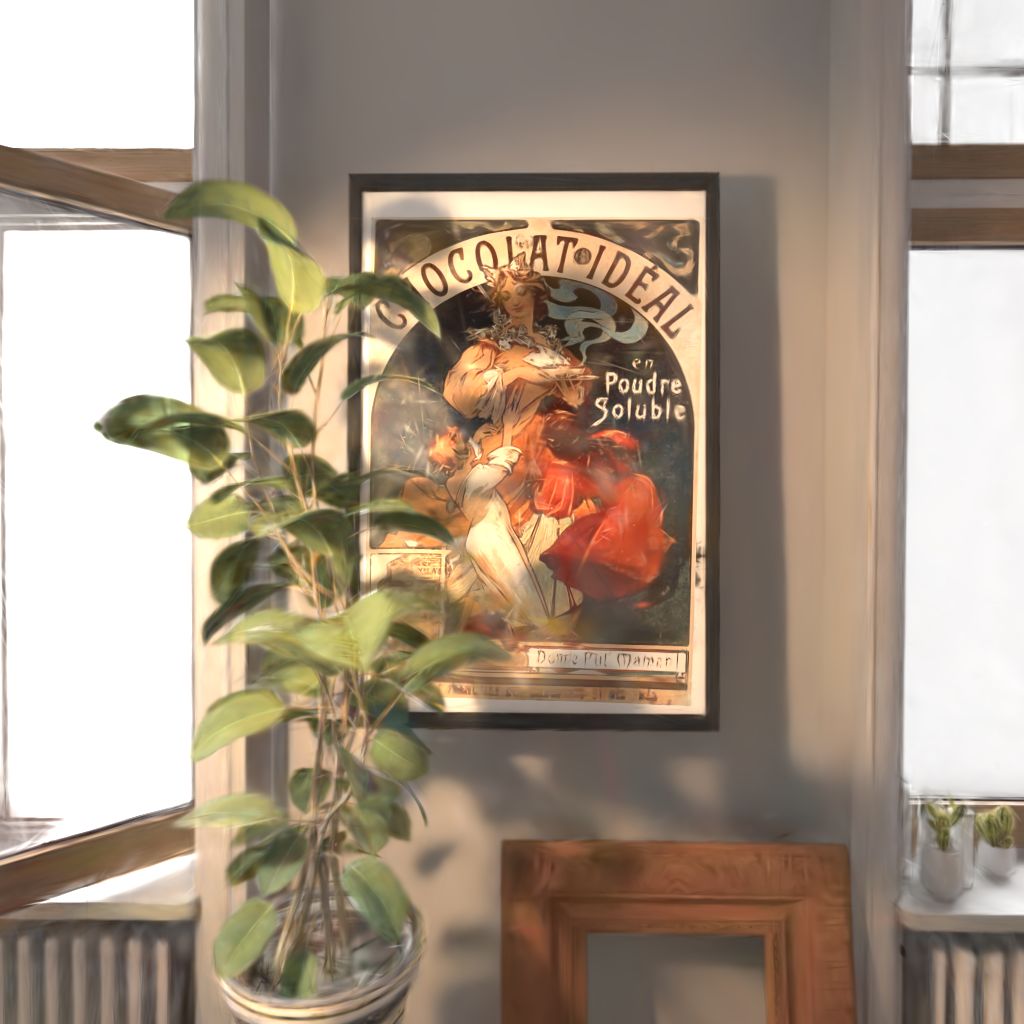} \\
    \hspace{-3mm}\footnotesize{Original Albedo} & \hspace{-3mm}\footnotesize{Edited Albedo} & \hspace{-3mm}\footnotesize{Shading} & \hspace{-3mm}\footnotesize{Re-rendering} \\
\end{tabular}
\caption{Intrinsic decomposition and albedo editing on a synthetic scene, using ground-truth albedo for supervision. By representing albedo and shading as two separate radiance fields, our method allows to introduce new visual details in the albedo without altering the soft shading and shadows.}
\label{fig_decomp_decal}
\vspace{4mm}
\end{figure}

We also show results on a synthetic scene that we use for the quantitative evaluation in the ablations (see Sec.~\ref{sec:ablations}). We rendered 288 images of this scene that we used as input photos for our intrinsic 3DGS pipeline.
Our results on this synthetic scene (Fig.~\ref{fig_decomp_decal}) illustrate the best possible outcome, since we use ground truth albedo and depth maps to reconstruct the intrinsic fields. 

\begin{table}[t]
\centering
\caption{Training time and primitive count across scenes.}
\label{tab:comparison_complexity}
\setlength{\tabcolsep}{4pt}
\begin{tabular}{l c c c c c c}
\toprule
Scene & Method & Layer & \# Primitives & ratio ($\times$ 3DGS) & Training Time & FPS\\
\midrule

\textbf{Basement 1} & Ours & Albedo & 1.575M & 1.01 & 00:06:56 & --\\
   & & Shading & 469K & 0.30 & 00:09:45 & --\\
   & & Residual & 141K & 0.09 & 00:11:49 & --\\
   & & Full & 2.185M & 1.40 & 00:28:30 & 96.33\\
\cmidrule(lr){2-7}
 & 3DGS & -- & 1.558M & 1 & 00:06:14 & 501.17\\
\midrule

\textbf{Basement 2} & Ours & Albedo & 851K & 0.40 & 00:04:28 & --\\
   & & Shading & 675K & 0.32 & 00:07:21 & --\\
   & & Residual & 70K & 0.03 & 00:09:02 & --\\
   & & Full & 1.596M & 0.76 & 00:20:51 & 103.52\\
\cmidrule(lr){2-7}
& 3DGS & -- & 2.113M & 1 & 00:06:19 & 448.95\\
\midrule

\textbf{Basement 3} & Ours & Albedo & 984K & 0.91 & 00:06:05 & --\\
   & & Shading & 272K & 0.25 & 00:08:56 & --\\
   & & Residual & 130K & 0.12 & 00:10:54 & --\\
   & & Full & 1.386M & 1.29 & 00:25:55 & 105.73\\
\cmidrule(lr){2-7}
& 3DGS & -- & 1.076M & 1 & 00:04:55 & 420.12\\
\midrule

\textbf{Synthetic room} & Ours & Albedo & 1.285M & 0.82 & 00:06:04 & --\\
   & & Shading & 752K & 0.48 & 00:09:44 & --\\
   & & Residual & 52K & 0.03 & 00:11:23 & --\\
   & & Full & 2.089M & 1.33 & 00:27:11 & 128.12\\
\cmidrule(lr){2-7}
& 3DGS & -- & 1.576M & 1 & 00:05:10 & 831.67\\
\bottomrule
\end{tabular}
\end{table}

\subsection{Comparisons}

\subsubsection{Intrinsic decomposition}
\begin{figure}[!h]
\centering
\begin{tabular}{ccccc}
    \rotatebox[origin=l]{90}{\hspace{15mm}\footnotesize{Albedo}} &
    \hspace{-3mm}\includegraphics[width=0.24\linewidth]{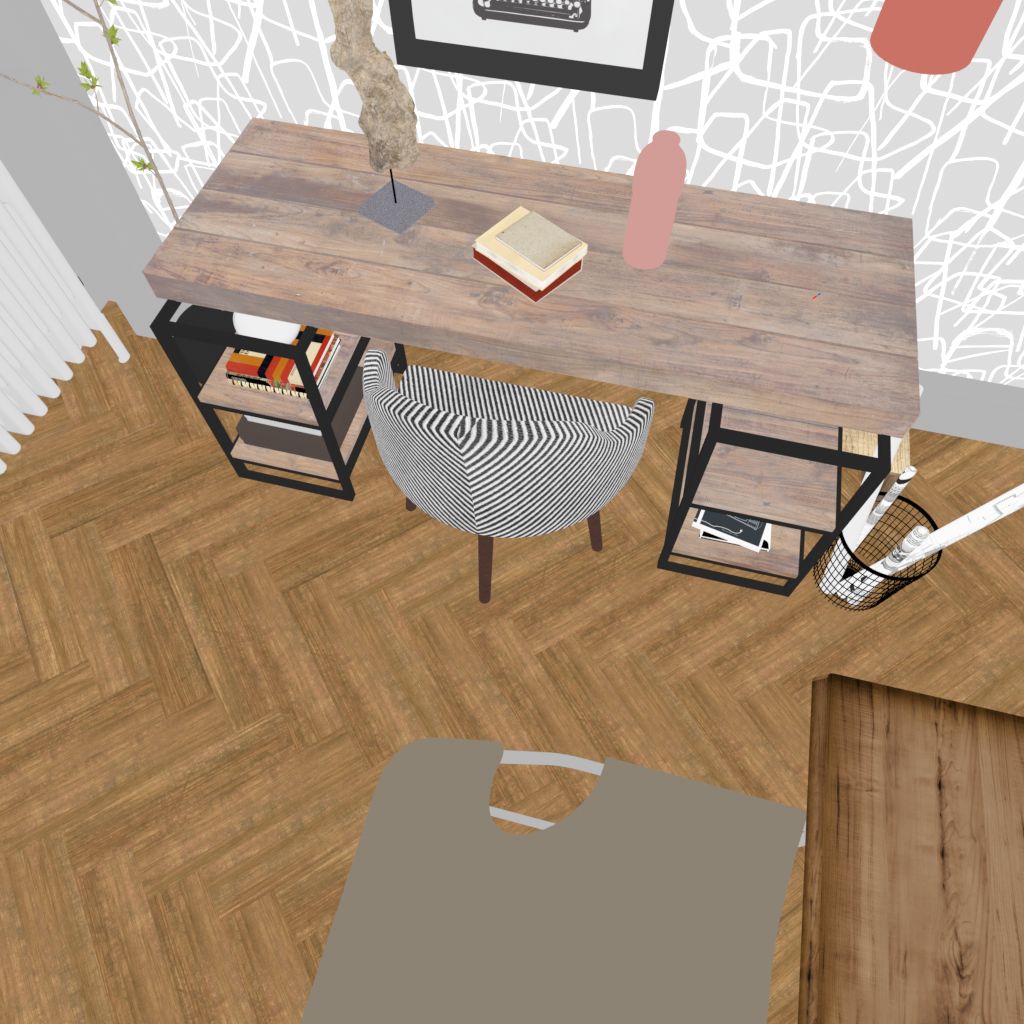} &
    \hspace{-3mm}\includegraphics[width=0.24\linewidth]{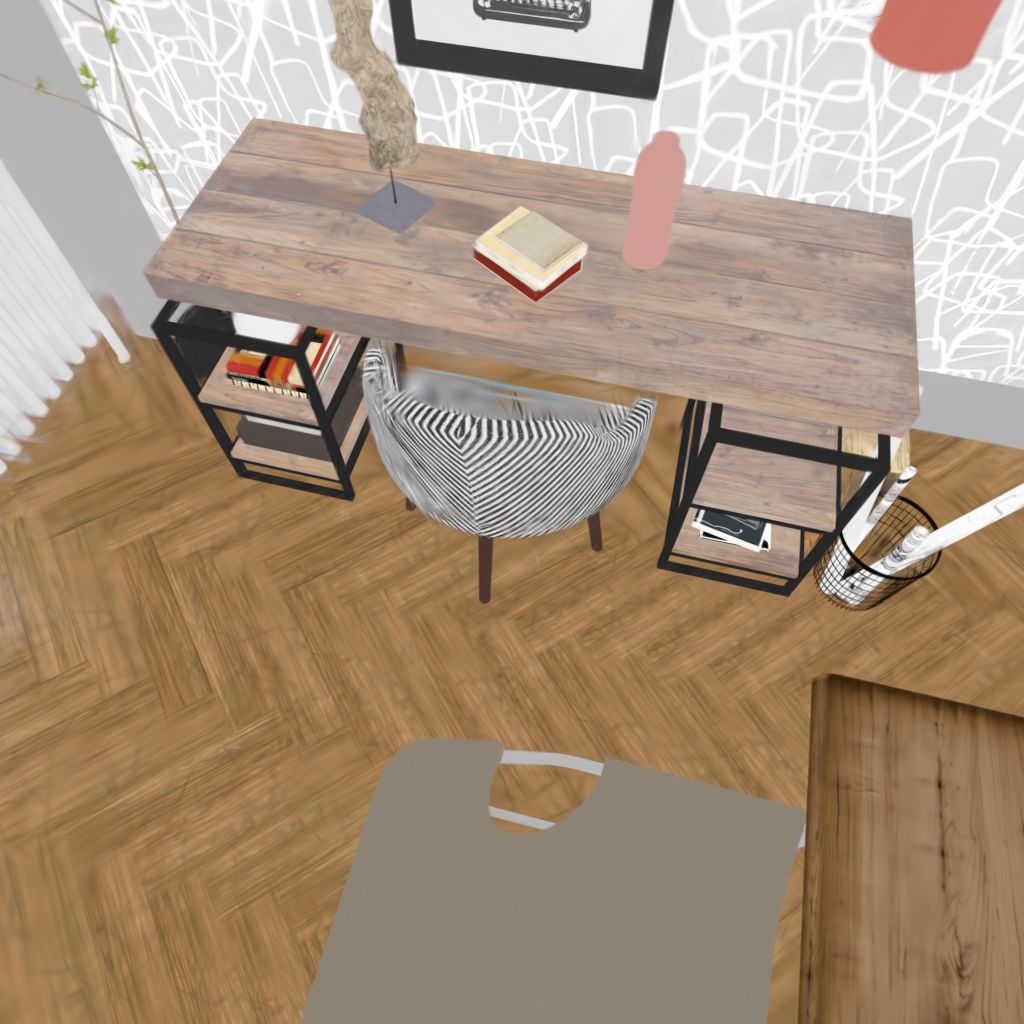} &
    \hspace{-3mm}\includegraphics[width=0.24\linewidth]{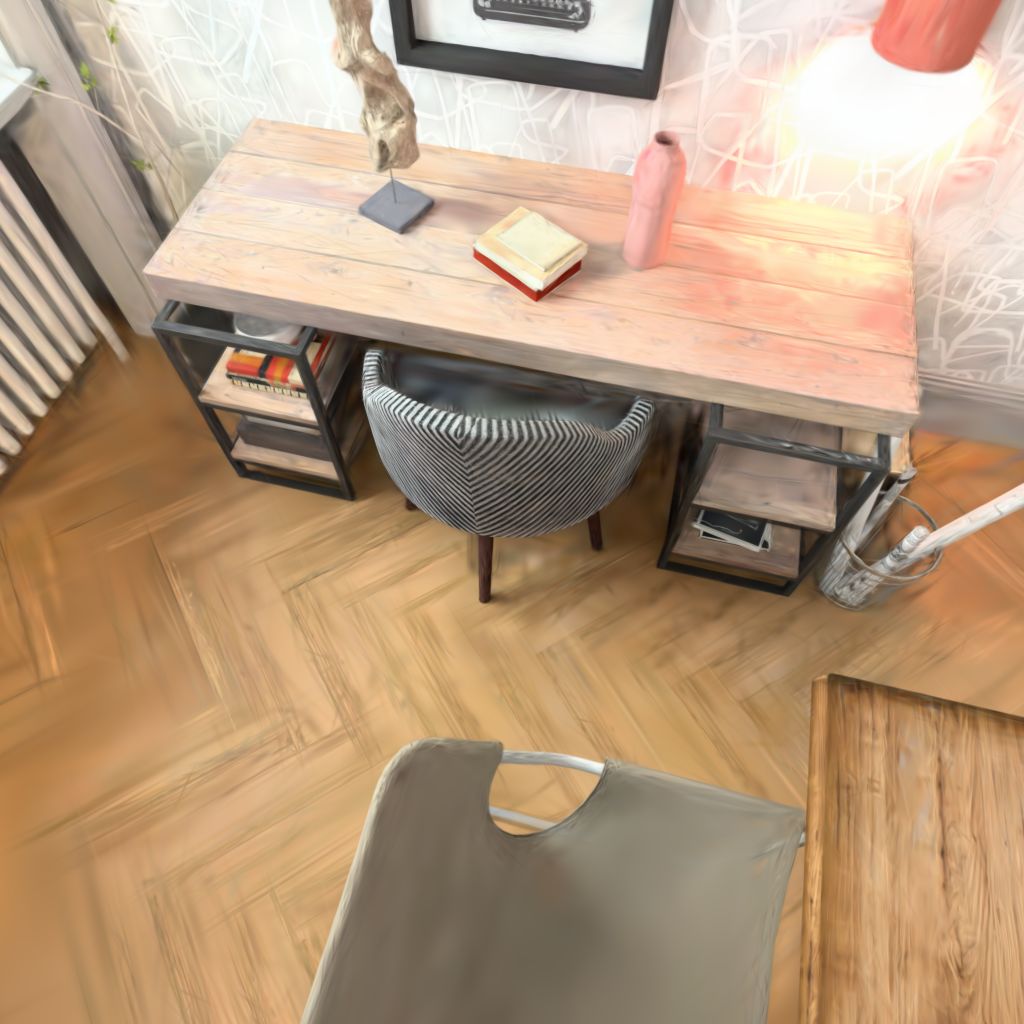} &
    \hspace{-3mm}\includegraphics[width=0.24\linewidth]{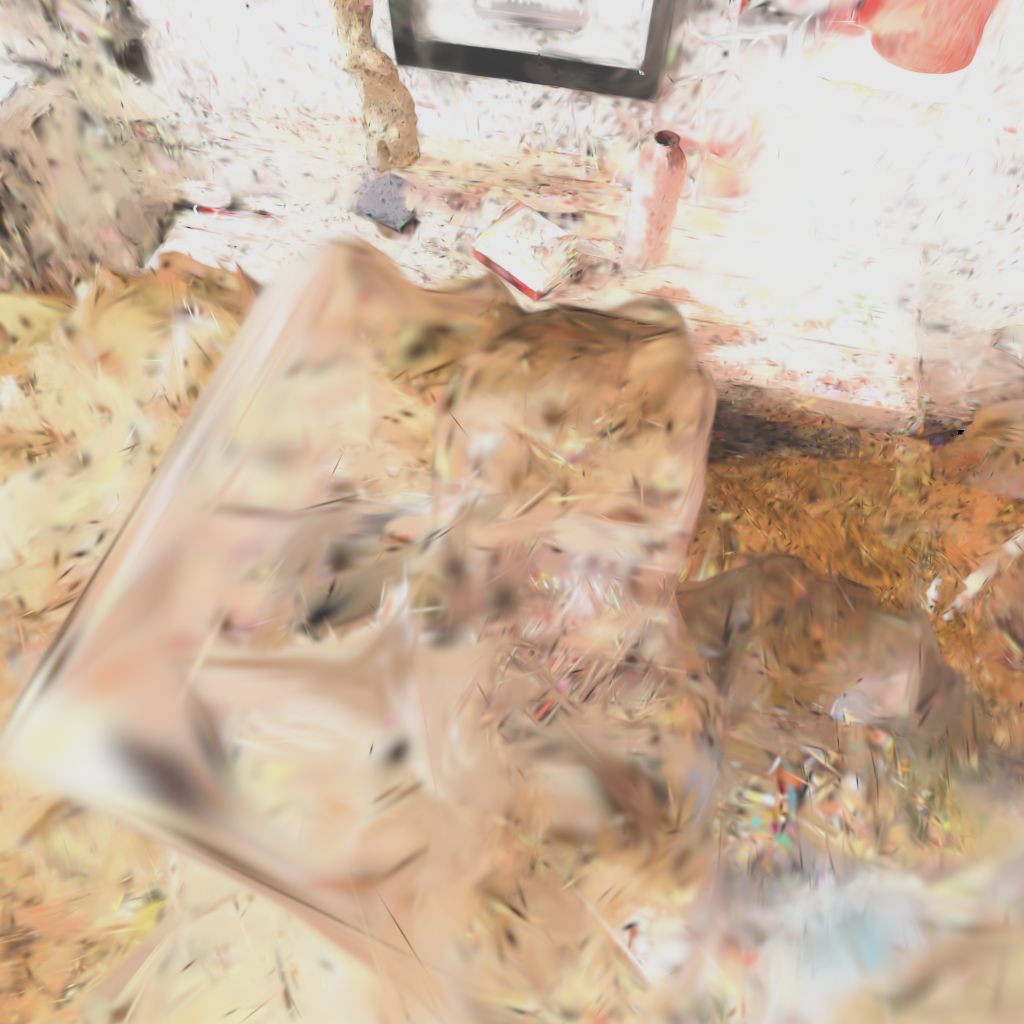} \\
    
    \rotatebox[origin=l]{90}{\hspace{15mm}\footnotesize{Full}} &
    \hspace{-3mm}\includegraphics[width=0.24\linewidth]{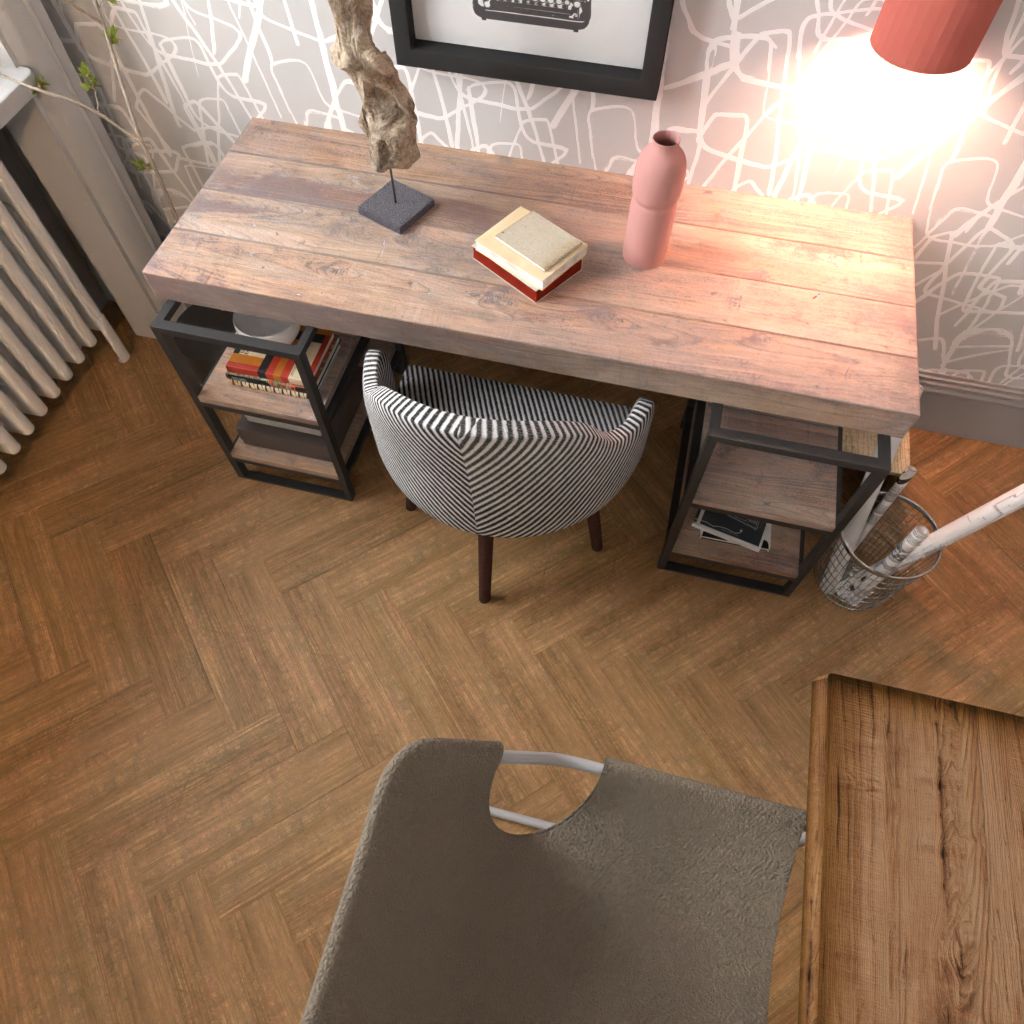} &
    \hspace{-3mm}\includegraphics[width=0.24\linewidth]{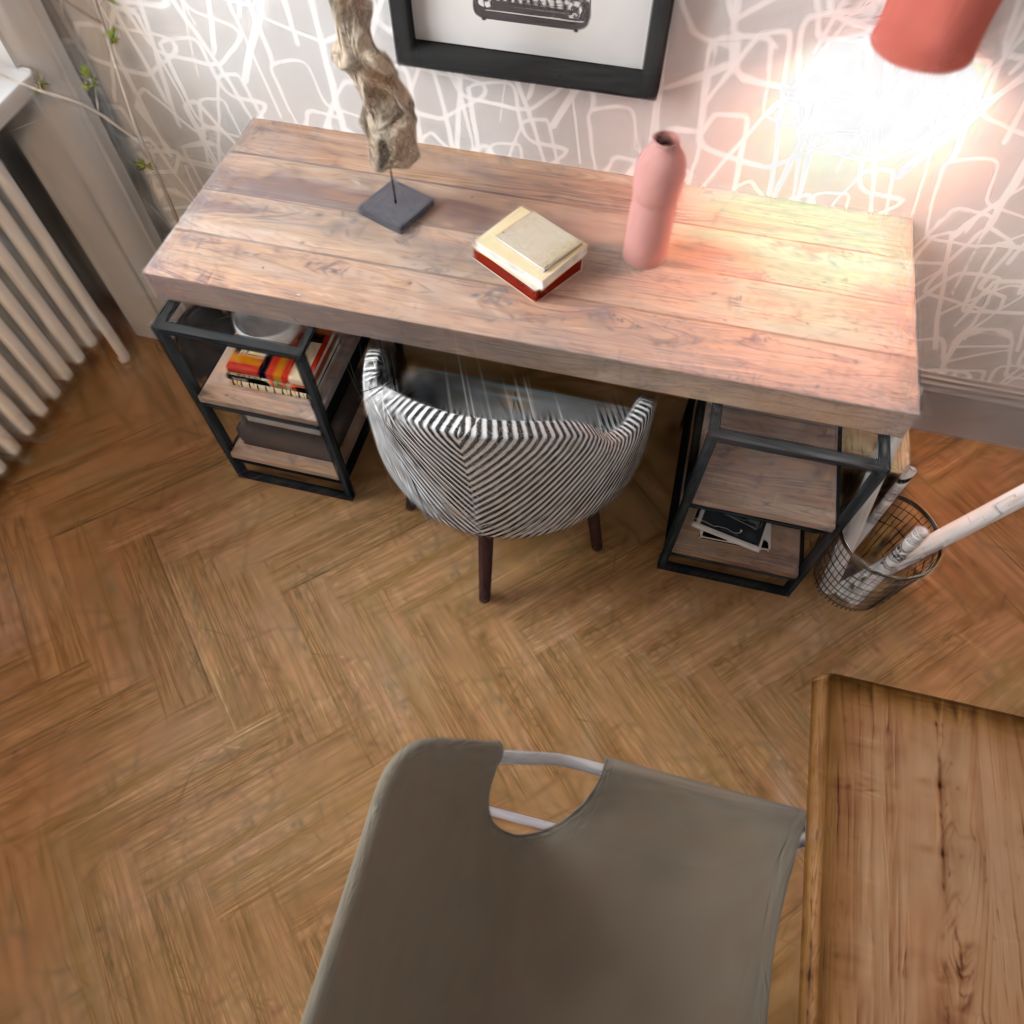} &
    \hspace{-3mm}\includegraphics[width=0.24\linewidth]{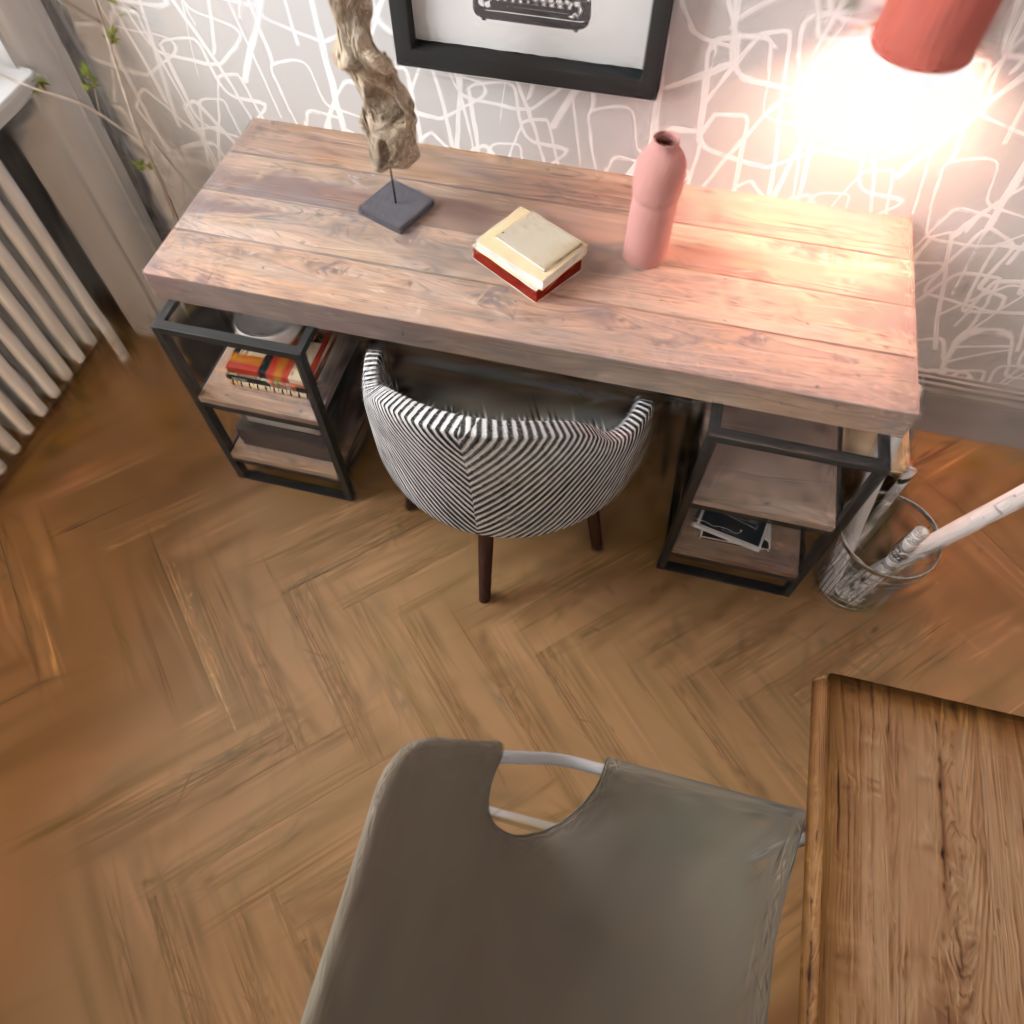} &
    \hspace{-3mm}\includegraphics[width=0.24\linewidth]{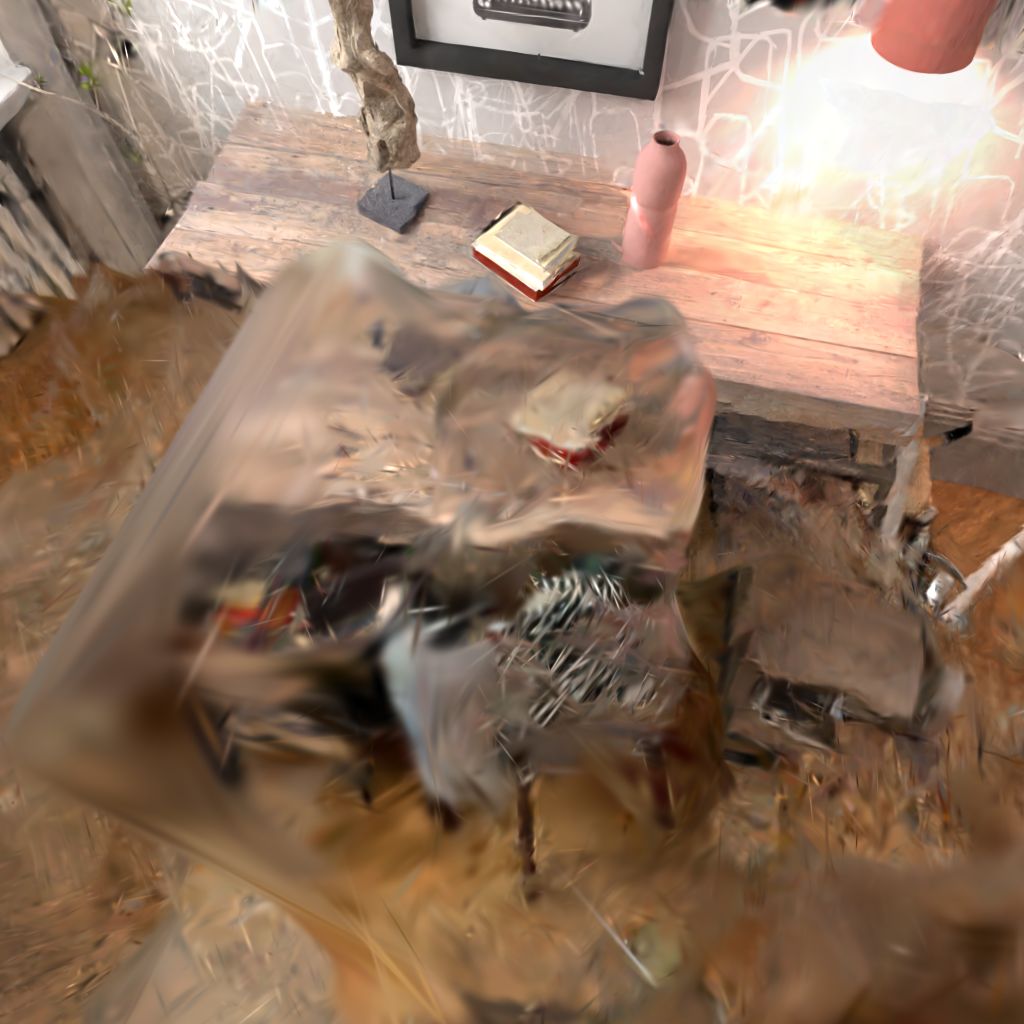} \\
    & \hspace{-3mm}\footnotesize{Ground truth} & \hspace{-3mm}\footnotesize{Ours} & \hspace{-3mm}\footnotesize{GI-GS} & \hspace{-3mm}\footnotesize{R3DG} 
\end{tabular}
\caption{
Compared to our approach, GI-GS \cite{chen2025gigs} retains significant shading residuals in the albedo, even though it better reconstructs the input image than R3DG \cite{R3DG2023}, which is designed for object-centric scenes and as such fails on our extended scene.}
\label{fig_comparison_intrinsics}
\end{figure}

We compare our intrinsic decomposition both quantitatively and qualitatively with two inverse rendering methods based on 3D Gaussian splatting on the synthetic scene shown as inset in Fig~\ref{fig:synthetic_scene}. R3GS~\cite{R3DG2023} and GI-GS~\cite{chen2025gigs} both propose a full inverse rendering pipeline using 3DGS, the former uses raytracing directly on Gaussians in 3D space while the latter relies on GBuffers obtained by rasterizing different properties on Gaussians to speed up the computations. Table.~\ref{tab:comparison_intrinsic} shows that on our synthetic indoor scene, our method produces higher quality albedo while being significantly faster than the others at the cost of generating $\times$3.5 more primitives than GI-GS~\cite{chen2025gigs}. R3DG~\cite{R3DG2023} also assumes a near natural white lighting so it cannot correctly represent the warmer, more realistic light placed in our synthetic scene; moreover it is designed for object-centric scenes so it fails to reconstruct our extended scene as shown in Fig~\ref{fig_comparison_intrinsics}.

\begin{table}[!t]
  \centering
  \caption{Quantitative results when comparing our intrinsic decomposition with GI-GS and R3GS.}
  \label{tab:comparison_intrinsic}
  \begin{tabular}{llcccccc}
    \toprule
    \textbf{Method} & \textbf{Layer} & \textbf{PSNR} & \textbf{SSIM} & \textbf{LPIPS} & \textbf{\# Primitives} & \textbf{Training Time}\\
    \midrule

    {Ours full}
      & Albedo & 29.419 &  0.932 & 0.095 & 1.285M & 00:06:04 \\
      & Full & 31.016 &  0.923 & 0.117 & 2.089M & 00:27:11 \\ %

    \midrule

    {GI-GS}
      & Albedo & 13.675 & 0.769 & 0.310 & -- & -- \\
      & Full & 35.284 & 0.951 & 0.106 & 790k & 00:50:49\\
    \midrule

    {R3GS}
      & Albedo & 11.680 & 0.683 & 0.445 & -- & -- \\
      & Full & 15.002 & 0.655 &  0.492 & 3.943M & 02:46:20 \\ %
    \bottomrule
  \end{tabular}
\end{table}

\subsubsection{Interactive edition}
We choose to compare our work on an interactive editing task against two other 3DGS editing methods that can perform recoloring. ReCoGS~\cite{rutayisire2025recogs} allows interactive tinting of Gaussians by reoptimizing the colors of the primitives affected by a user-specified mask. The second method we compare to is SplatShop~\cite{schutz2025splatshop}, which provides a framework to select, transform and directly paint Gaussians that are intersected with a 3D brush. SplatShop is tailored for VR and the painting does not involve any optimization.

\begin{figure}[!h]
\centering
\begin{tabular}{cccccc}
    \rotatebox[origin=l]{90}{\hspace{6mm}\footnotesize{Synthetic}} &
    \hspace{-3mm}\includegraphics[width=0.19\linewidth]{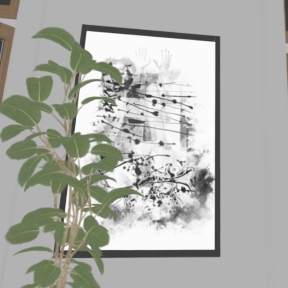} &
    \hspace{-3mm}\includegraphics[width=0.19\linewidth]{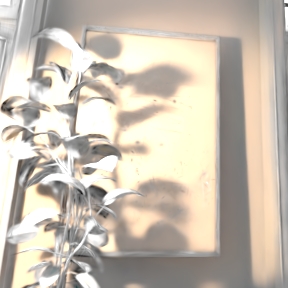} &
    \hspace{-3mm}\includegraphics[width=0.19\linewidth]{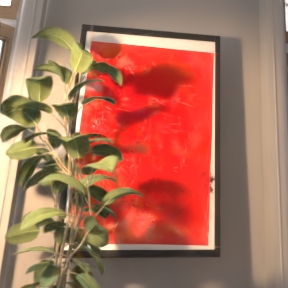} &
    \hspace{-3mm}\includegraphics[width=0.19\linewidth]{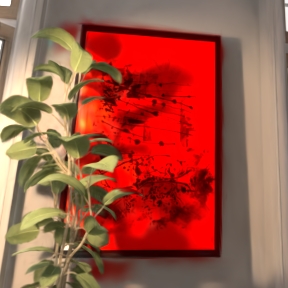} &
    \hspace{-3mm}\includegraphics[width=0.19\linewidth]{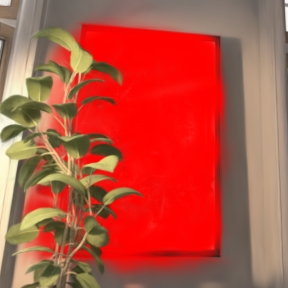} \\
    \rotatebox[origin=l]{90}{\hspace{15mm}\footnotesize{Real}} &
    \hspace{-3mm}\includegraphics[width=0.19\linewidth]{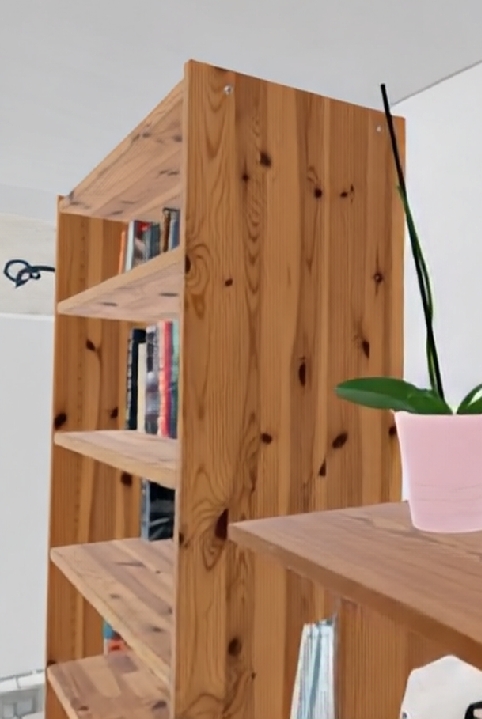} &
    \hspace{-3mm}\includegraphics[width=0.19\linewidth]{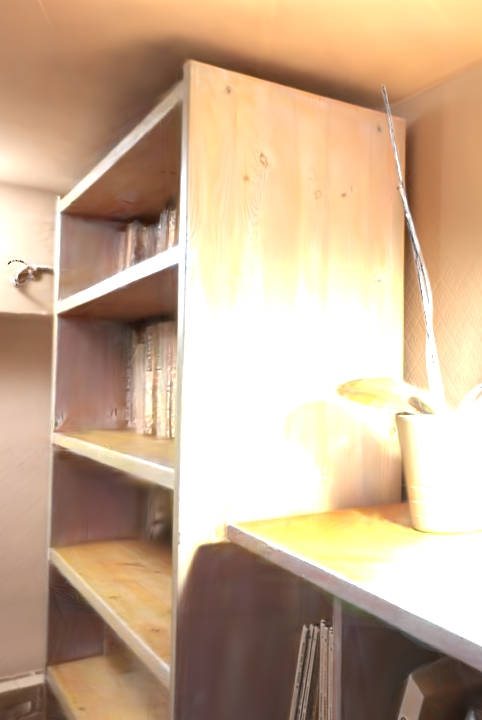} &
    \hspace{-3mm}\includegraphics[width=0.19\linewidth]{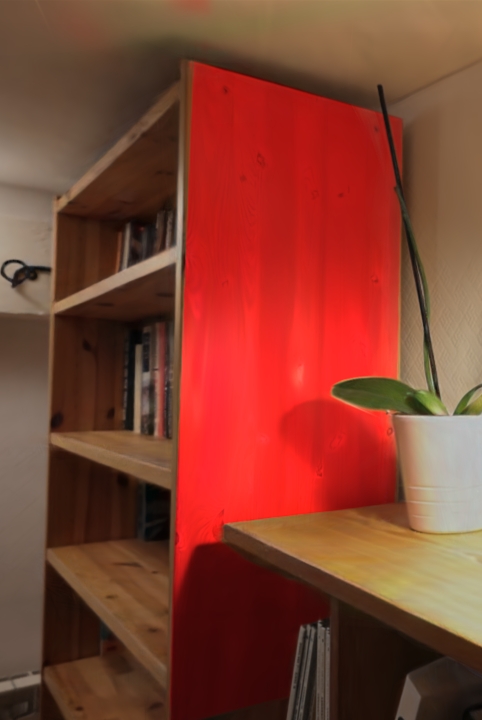} &
    \hspace{-3mm}\includegraphics[width=0.19\linewidth]{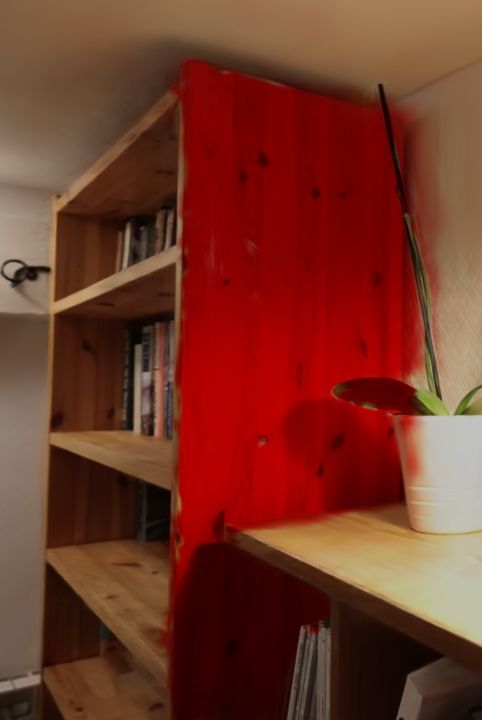} &
    \hspace{-3mm}\includegraphics[width=0.19\linewidth]{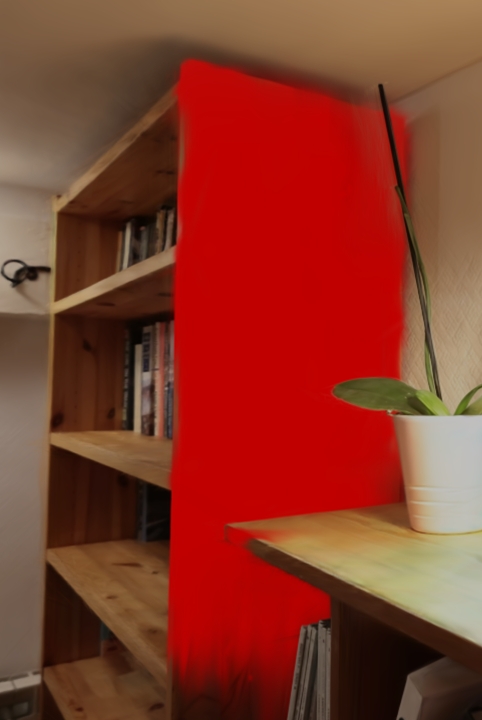} \\
    & \hspace{-3mm}\footnotesize{Albedo} & \hspace{-3mm}\footnotesize{Shading} & \hspace{-3mm}\footnotesize{Ours} & \hspace{-3mm}\footnotesize{ReCoGS} & \hspace{-3mm}\footnotesize{SplatShop}
\end{tabular}
\caption{
Our intrinsic decomposition enables realistic editing of colors and textures compared to previous methods where shading and albedo are entangled. Our method correctly preserves shadows while replacing the texture, while ReCoGS incorrectly preserves the texture \cite{rutayisire2025recogs}, and SplatShop \cite{schutz2025splatshop} overwrites both texture and shadows. Note however that slight texture details remain visible in our result for the real scene (bottom) due to the albedo predictor that tends to produce blurry albedo maps. }
\label{fig_comparison_decomp}

\end{figure}

This recoloring application showcases two advantages of our method compared to previous work, as it allows to modify the albedo without altering the shading. This is illustrated in Fig.~\ref{fig_comparison_decomp} where we recolored textured surfaces with a uniform red color, for a synthetic scene and a real scene. 
Setting a red tint in ReCoGS preserves both the shading and the texture information, while painting with a red brush in SplatShop erases both the shading and the texture. In contrast, our method allows to remove the texture without altering the shading and cast shadows. Note however that on the real scene, errors in the albedo prediction makes some albedo details appear in the shading, and as such remain visible in our edit, albeit much less than with ReCoGS. Our decomposition also yields a sharper boundary around the edited area, which we achieve by updating the parameters of the albedo Gaussians while keeping the shading and residual Gaussians untouched.

\subsection{Ablation Study}
\label{sec:ablations}

We evaluate our method across different configurations on the quality of the reconstructed albedo field, shading field and fully composited renderings using the image formation model of Eq.~\ref{eq:basic_im}.

We use the synthetic scene shown as inset to perform quantitative evaluation, 
comparing the rendered results of each reconstructed field with the values of 
ground truth renderings. %

\begin{figure}
   \centering
   \begin{tabular}{cc}

	   \includegraphics[width=0.45\linewidth]{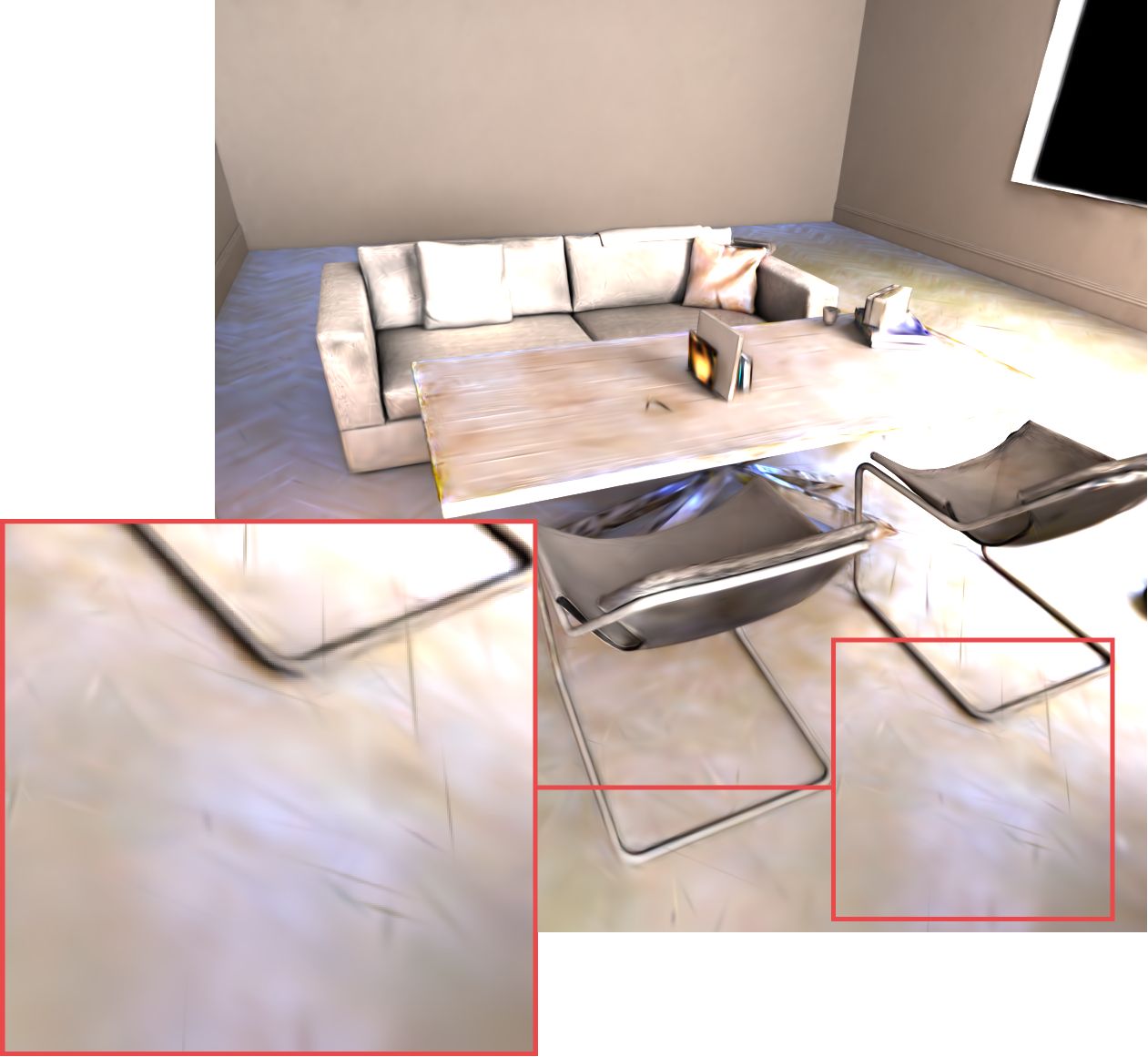} &
       \includegraphics[width=0.45\linewidth]{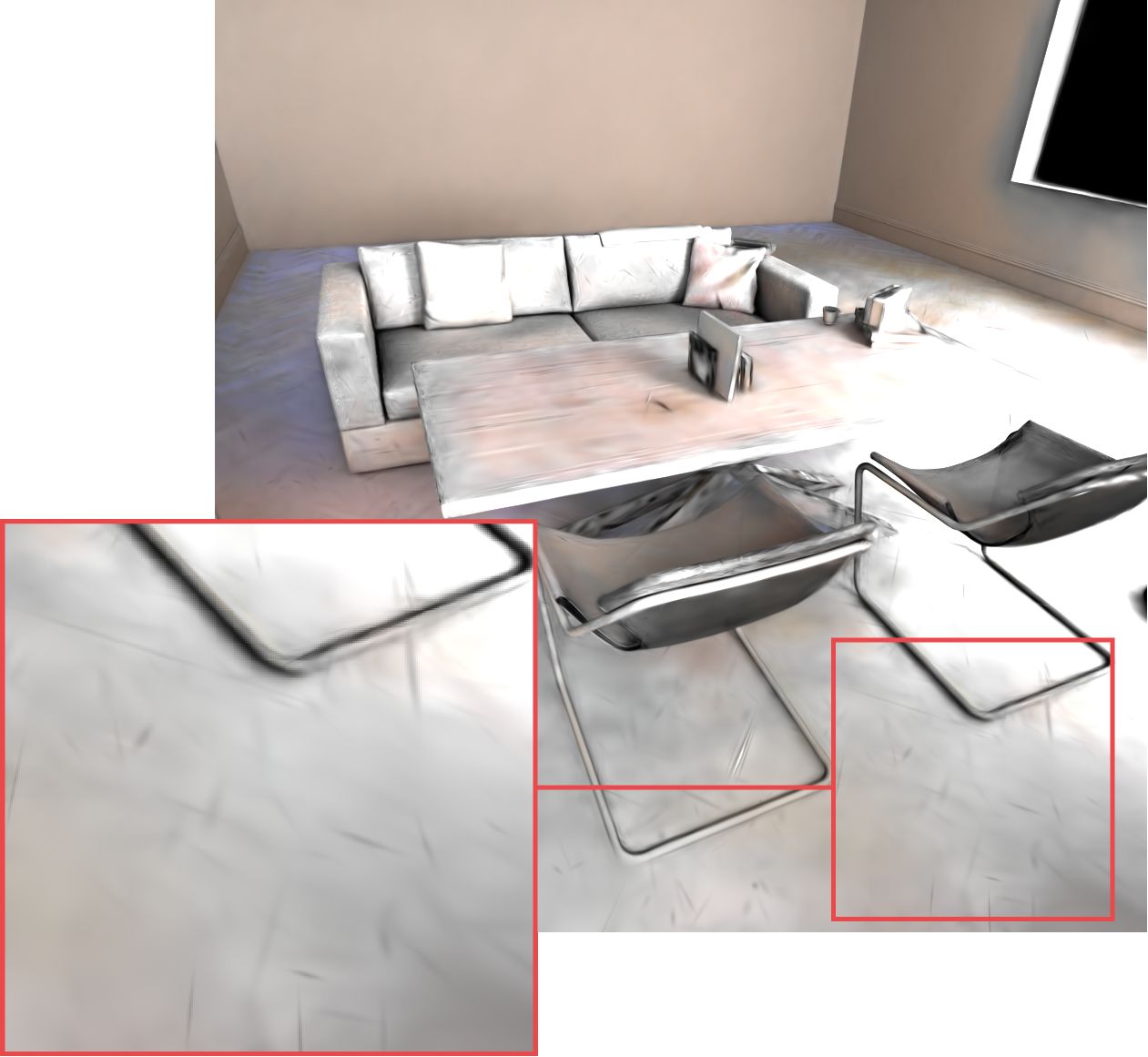}  \\

	   \footnotesize{Without color regularization} & \footnotesize{Ours} \\
   \end{tabular}
   \caption{Shading field reconstruction with color regularization turned off/on. Our shading reconstruction is more constrained while remaining colorful.}
   \label{fig:ablation_shading_cr}
\end{figure}

\noindent
\begin{minipage}[t]{0.65\textwidth} %
  To compute our error metrics for the ablations, we left out every 8th image 
  of the dataset as test views and compute PSNR, SSIM and LPIPS on albedo, 
  shading and fully composited renderings. The results of the quantitative evaluation are shown in Tab.~\ref{tab:metrics}. 
  
  Our color regularization (Eq.~\ref{eq:col_reg}) yields a cleaner shading reconstruction that contains fewer artifacts while remaining colorful; this regularization does not degrade the reconstruction quality, as can be seen in the metrics that are essentially unchanged. The qualitative effect of the regularization can be seen in Fig.~\ref{fig:ablation_shading_cr}.
\end{minipage}
\hfill
\begin{minipage}[t]{0.30\textwidth} %
  \centering
  \vspace{0pt} %
  \includegraphics[width=\linewidth]{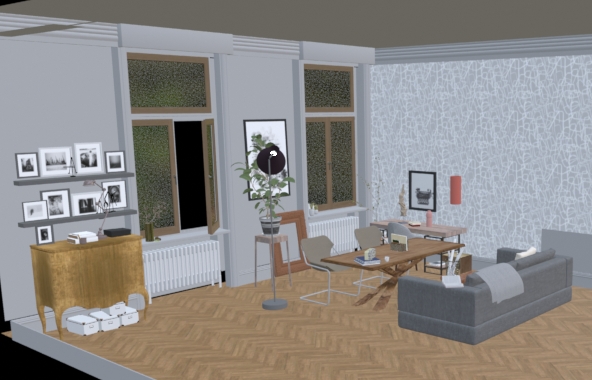}
    \vspace{-20pt} %
    \captionof{figure}{Our synthetic scene}
  \label{fig:synthetic_scene}
  \small
  \vspace{4pt}
\end{minipage}

\noindent

\begin{table}[!t]
  \centering
  \caption{Quantitative results of ablations, namely 
  shading reconstruction without Color Regularization (CR*) and albedo reconstruction using predicted albedo.}
  \label{tab:metrics}
  \begin{tabular}{llccc}
    \toprule
    \textbf{Method} & \textbf{Layer} & \textbf{PSNR} & \textbf{SSIM} & \textbf{LPIPS} \\
    \midrule

    {Ours full}
      & Albedo & 31.827 &	0.945 &	0.078 \\
      & Shading & 22.876 & 0.894 & 0.210 \\
      & Full & 35.679 & 0.956 & 0.087 \\
    \midrule

    {Shading w/o CR*}
      & Shading & 22.957 & 0.897 & 0.208\\
    \midrule

    {Predicted albedo}
      & Shading & 14.114 & 0.744 & 0.295 \\
      & Full & 34.278 & 0.944 & 0.126 \\
    \bottomrule
  \end{tabular}
\end{table}

\begin{figure}
  \centering
  \begin{tabular}{cc}
    \includegraphics[width=0.45\linewidth]{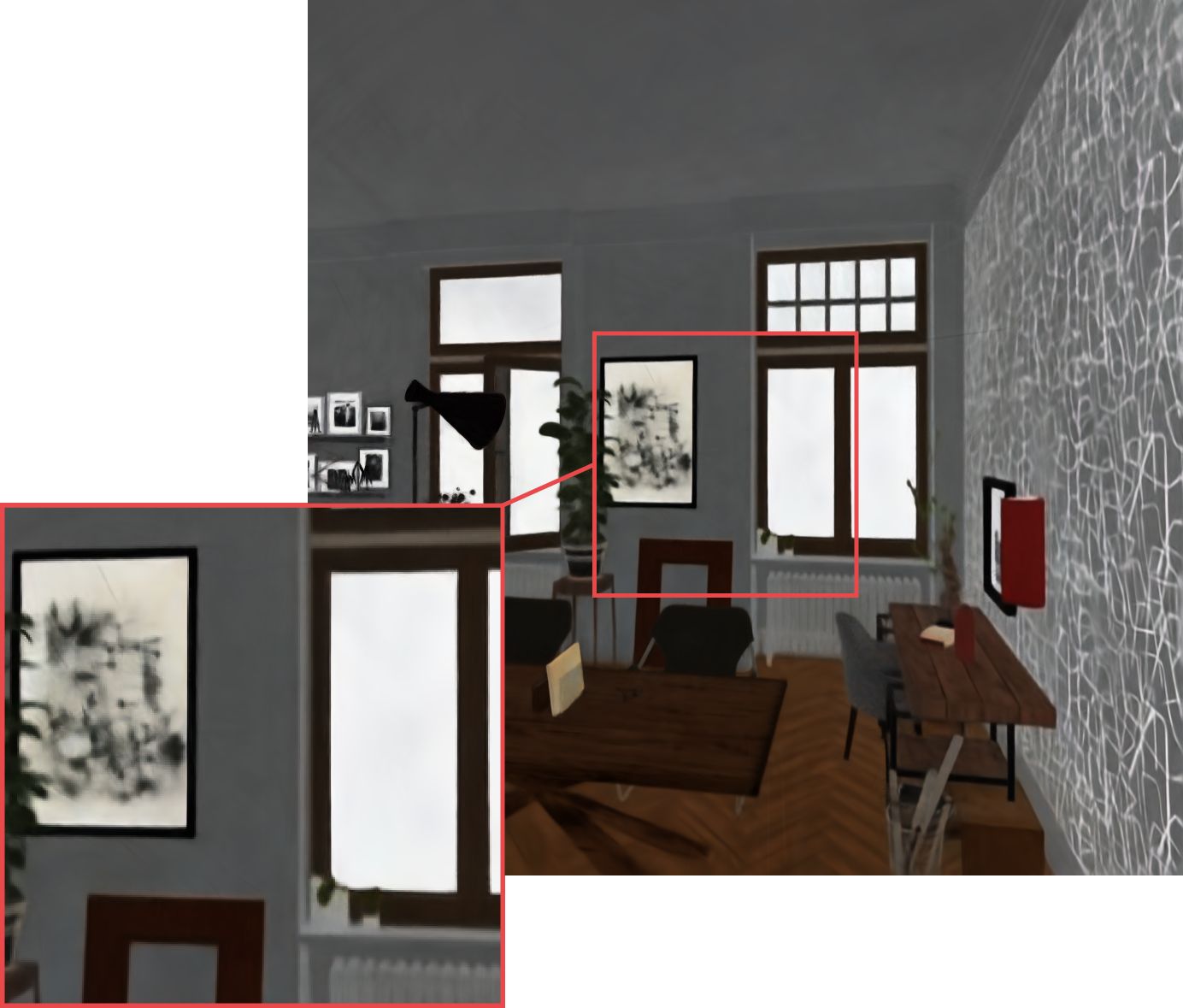} &
    \includegraphics[width=0.45\linewidth]{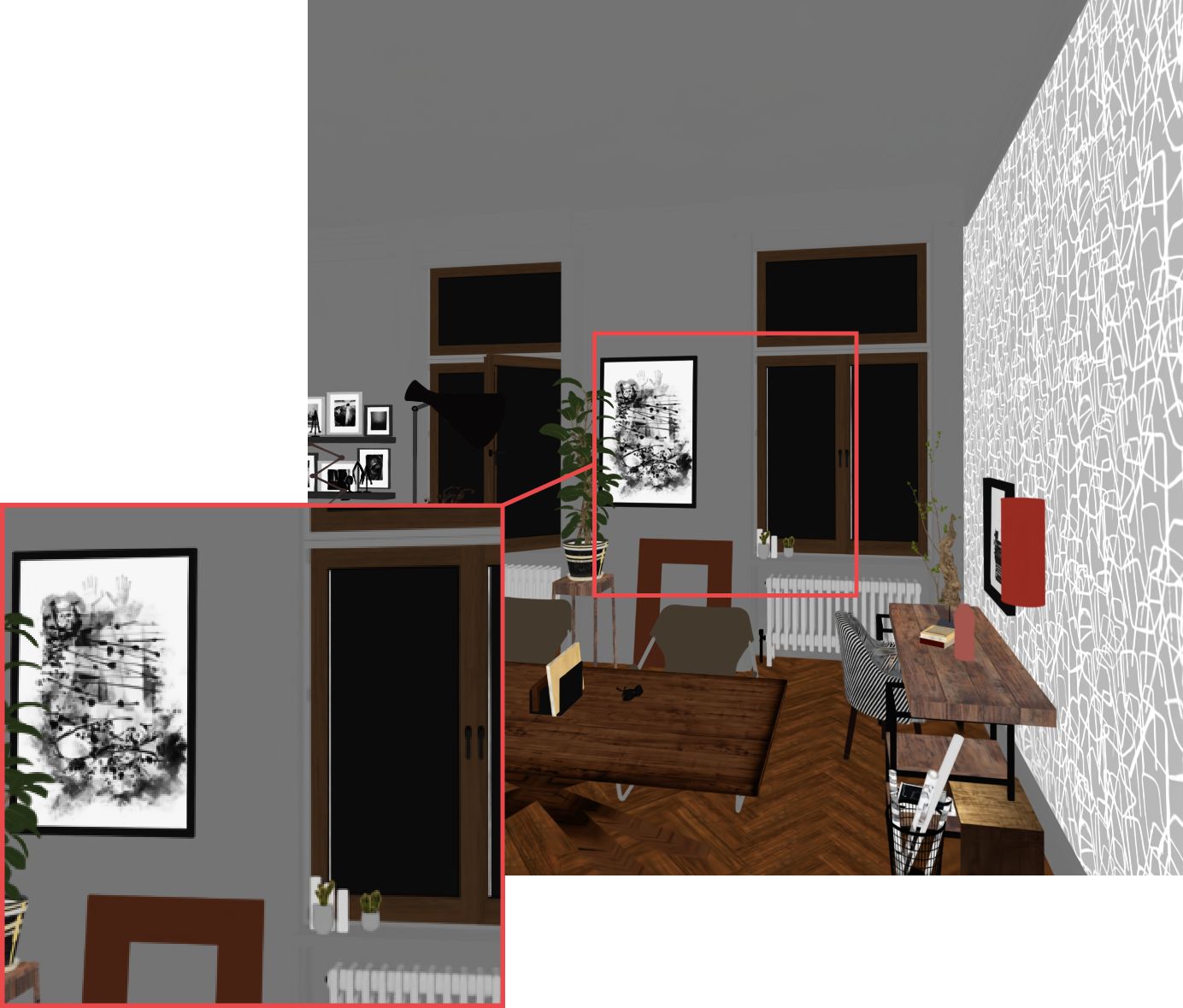} \\

	  \footnotesize{With predicted albedo} & \footnotesize{With ground-truth albedo} \\
  \end{tabular}
  \caption{Predicted albedo images can lack details and exhibit a slight color shift due to unknown nonlinearities, which yields blurrier reconstructions of the albedo field (left) compared to the reconstruction obtained with exact albedo images (right).}
  \label{fig:ablation_albedo_pred}
\end{figure}

Replacing ground truth albedo maps with predicted albedo maps gives a blurrier albedo reconstruction, which makes it harder to reconstruct shading and residual layers free of albedo variations (Fig.~\ref{fig:ablation_albedo_pred}). We tried several other options, but the multi-view consistent albedo maps from DiffusionRenderer are currently the best option, even if quite blurry. Moreover, we observed that DiffusionRenderer does not produce linear outputs. While the gamma inversion we apply compensates for this nonlinearity, it is only an approximation and some color shift sometimes remains.

\section{Limitations}

While our method allows a reasonable intrinsic decomposition of 3D Gaussian Splatting, it comes with some limitations.

\begin{figure*}
\centering
\begin{tabular}{cccc}
     \hspace{-3mm}\includegraphics[width=0.25\linewidth]{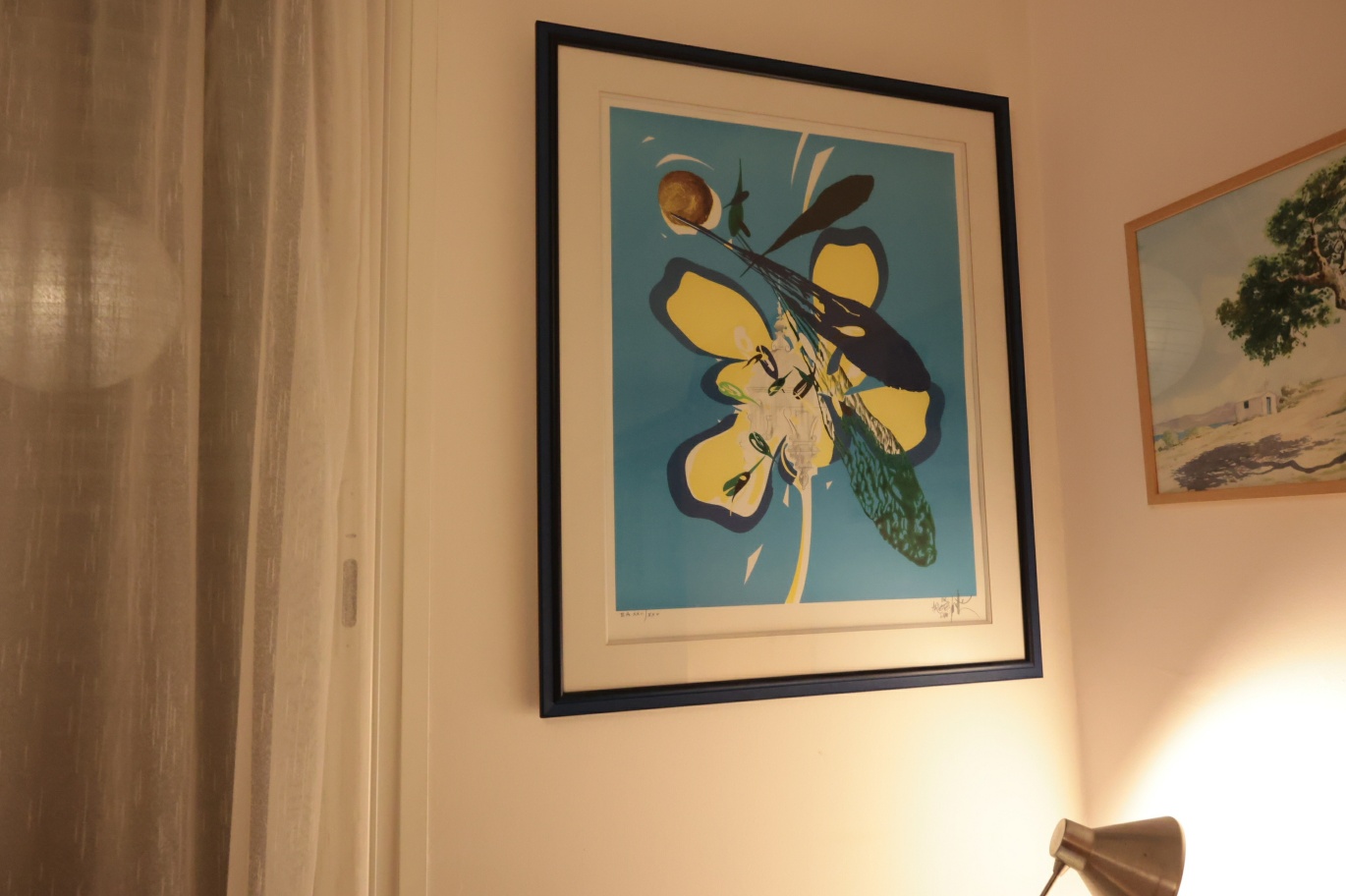} &
     \hspace{-3mm}\includegraphics[width=0.25\linewidth]{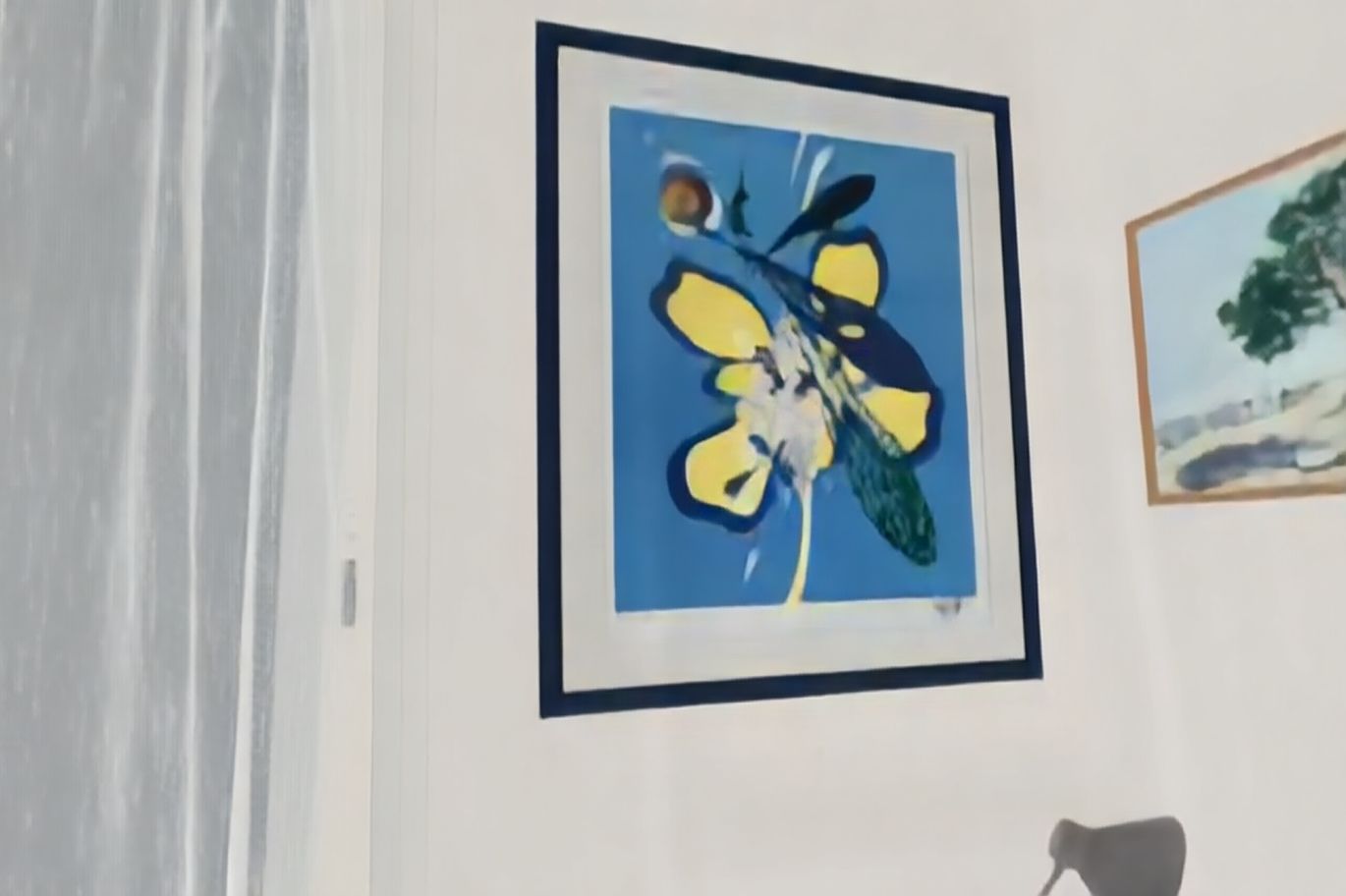} &
     \hspace{-3mm}\includegraphics[width=0.25\linewidth]{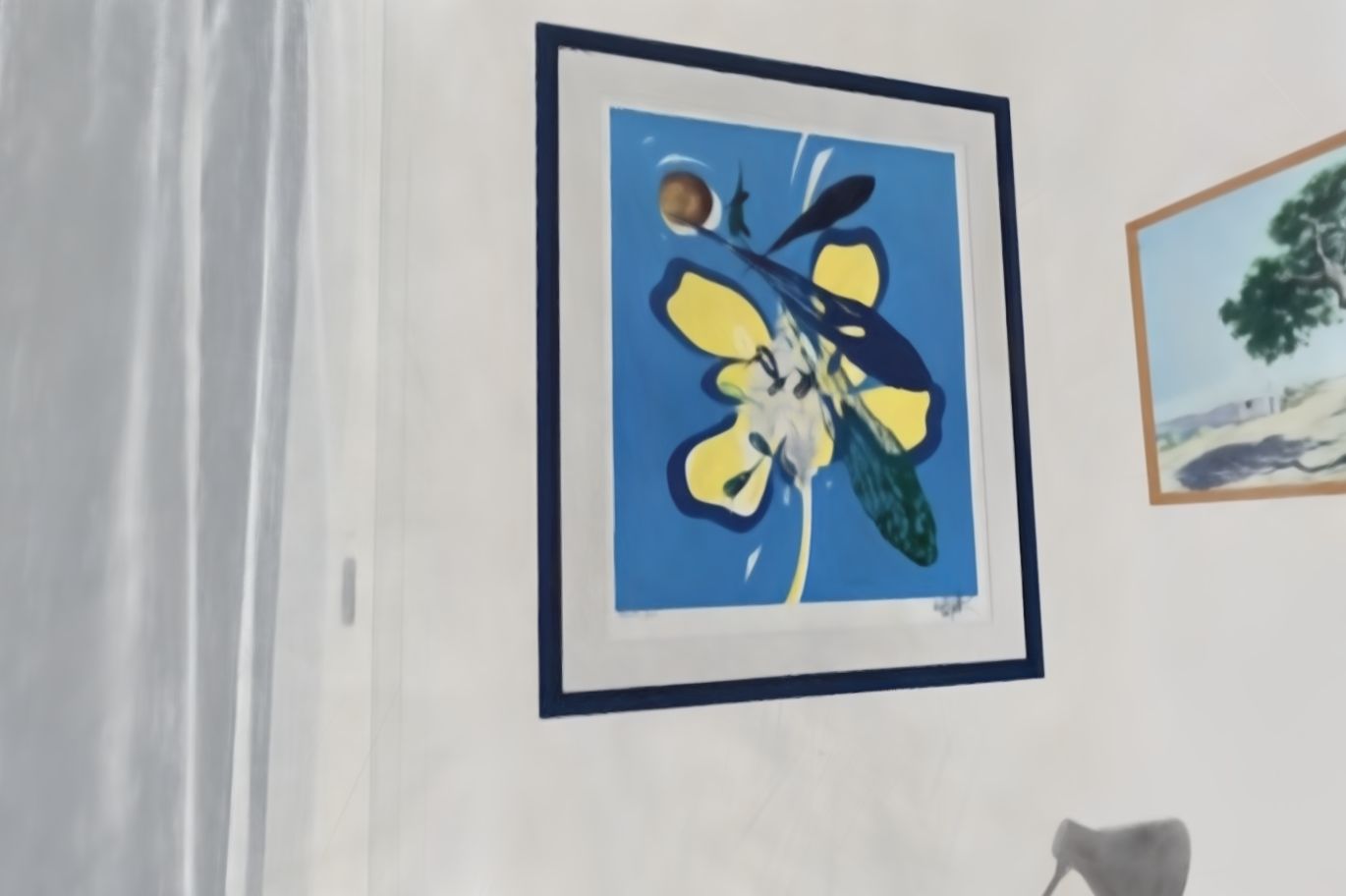} &
     \hspace{-3mm}\includegraphics[width=0.25\linewidth]{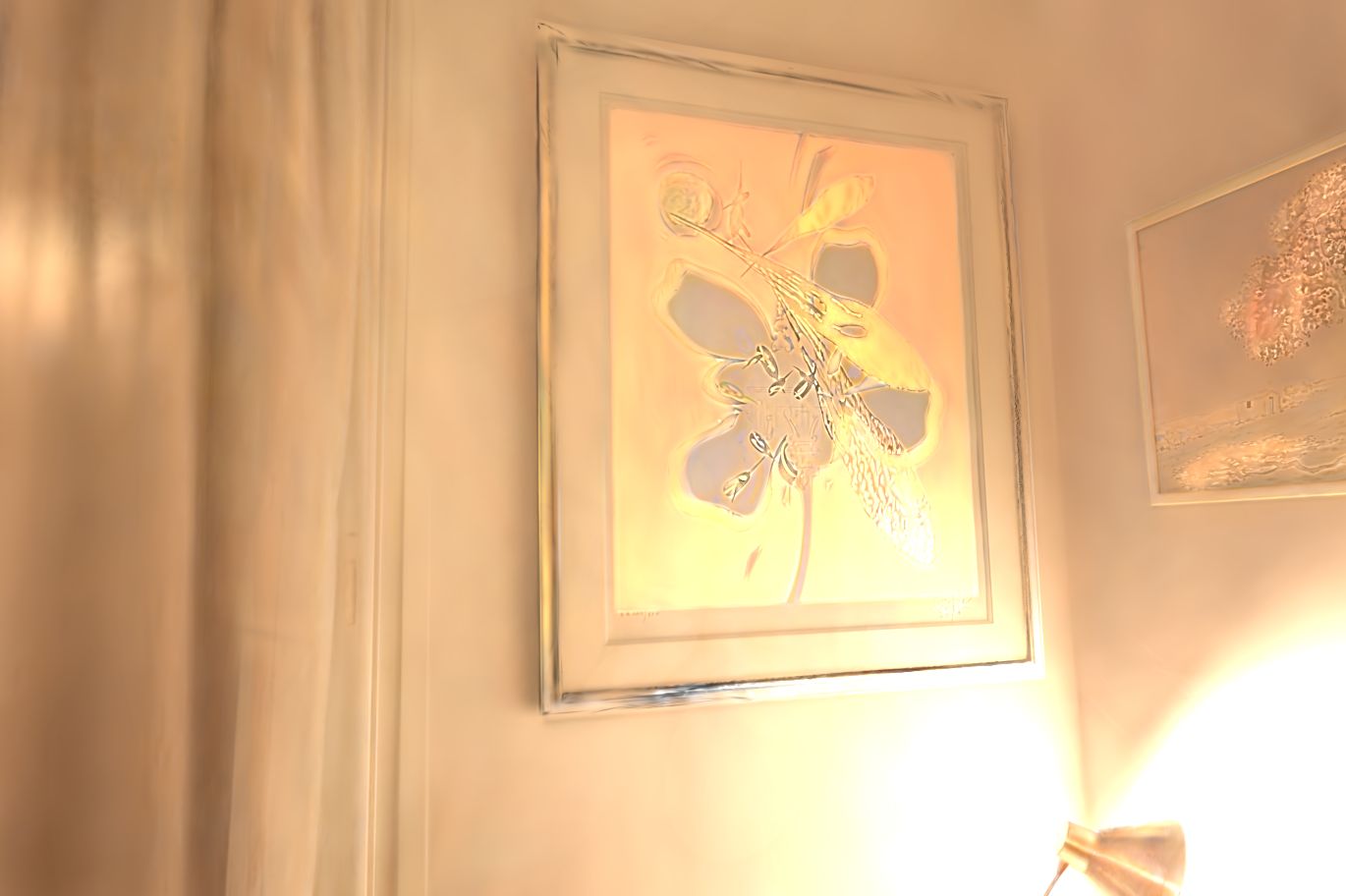} \\
     \hspace{-3mm}\footnotesize{(a) Input image} & \hspace{-3mm}\footnotesize{(b) Predicted Albedo} & \hspace{-3mm}\footnotesize{(c) Rendered Albedo} & \hspace{-3mm}\footnotesize{(d) Rendered Shading}
\end{tabular}
\caption{
\label{fig:limitation}
DiffusionRenderer tends to predict blurry albedo images (b). As a result, the high-frequency details present in the original images (a) are reconstructed by the shading field (d).
}
\end{figure*}

The first limitation is the ``fuzzy'' nature of the 3DGS representation itself, that makes editing complicated. We have demonstrated a practical solution for edits on planar surfaces, requiring some user interaction. This could be extended to simple non-planar shapes by fitting suitable proxies (e.g., cylinders or surfaces of revolution), but providing a truly general solution would be more complex. Methods that compute meshes from 3D Gaussian splats (e.g., \cite{guedon2023sugar,guedon2025milo}) could potentially be used to initialize proxies, but accurate edits would still be challenging. Our method also inherits popping artifacts already present in standard 3DGS reconstruction, most notably seen in darker regions. These could be reduced by sorting the primitives as in StopThePop \cite{radl2024stopthepop}.
The second limitation is our dependence on the quality of diffusion models to extract albedo layers. While the layers can be used in a useful manner as shown in our results on real scenes, the predictions are often blurry, and sometimes can assign texture to shading. An example is shown in Fig.~\ref{fig:limitation}. Future progress in intrinsic image and video decomposition will directly benefit our method. Finally, as previously mentioned in section \ref{sec:results} our method is slower than 3DGS but a caching mechanism could be implemented to avoid re-rendering  pretrained layers when training. At inference time this caching mechanism cannot be used but smarter memory layout of the Gaussians representing different layers could be used to increase rendering speed, since the current implementation consists of calling the renderer 3 times with different sets of Gaussians.

\section{Future Work and Conclusion}

In future work, we would like to investigate how to build on our ideas to allow full inverse rendering for radiance fields. Our explicit decomposition is promising, but many challenges remain in terms of the actual representation and the capacity to perform light transport.

We have presented a new method for intrinsic decomposition of 3D Gaussian splats, using diffusion model priors for albedo and depth. We introduce a new approach that first predicts an \emph{albedo field}, and then jointly optimizes a \emph{shading} and \emph{residual} field. This intrinsic decomposition of 3D Gaussian splats allows easy editing of albedo and texture, while maintaining the appearance of shading, allowing for realistic edited results.

\begin{acks}
This work was funded by the European Research Council (ERC) Advanced Grant NERPHYS, number 101141721 https://project.inria.fr/nerphys.  Views and opinions expressed are however those of the author(s) only and do not necessarily reflect those of the EU or the European Research Council. Neither the EU nor the granting authority can be held responsible for them. The authors are grateful to the OPAL infrastructure of the Université Côte d’Azur for providing resources and support, as well as Adobe and NVIDIA for software and hardware donations. Experiments presented in this paper were carried out using the Grid'5000 testbed, supported by a scientific interest group hosted by Inria and including CNRS, RENATER and several Universities as well as other organizations (see https://www.grid5000.fr).     
\end{acks}

\clearpage
\bibliographystyle{ACM-Reference-Format}
\bibliography{intrinsic_radiance.bib}

\end{document}